\documentclass{aa}  

\usepackage{graphicx}
\usepackage{txfonts}
\usepackage{aas_macros}
\usepackage{color}			
\usepackage{breakurl}		

\newcommand{\given}{\,\vert\,}

\begin{document}
\title{Time-dependent properties of sunspot groups -- I. Lifetime and asymmetric evolution}

   \author{
   	  Emese Forgács-Dajka
          \inst{1}
          \and
          László Dobos
          \inst{2}
          \and
          István Ballai
          \inst{3}
          }

   \institute{Department of Astronomy, E\"otv\"os Lor\'and University, H-1117 Budapest, P\'azm\'any P\'eter s\'et\'any 1/A, Hungary \\
              \email{e.forgacs-dajka@astro.elte.hu}
         \and
             Department of Physics \& Astronomy, The Johns Hopkins University, 3400 North Charles Street, Baltimore, MD 21218, USA \\
             Department of Physics of Complex Systems, E\"otv\"os Lor\'and University, H-1117 Budapest, P\'azm\'any P\'eter s\'et\'any 1/A, Hungary \\
             \email{dobos@jhu.edu}
         \and
             Plasma Dynamics Group, University of Sheffield, Hounsfield Road, Hicks Building, Sheffield, S3 7RH, United Kingdom \\
             \email{i.ballai@sheffield.ac.uk}
             }

   \date{Received ; accepted }

 
 \abstract
   {}
   {In this paper, we aim to study the time dependence of sunspot group areas in a large sample composed of various databases spanning over 130 years, used state-of-the-art statistical methods.}
   {For a carefully selected but unbiased sample, we use Bayesian modelling to fit the temporal evolution of the combined umbral and penumbral area of spot groups with a skew-normal function to determine the existence of any asymmetry in spot growth or decay. Our primary selection criteria guaranteed that only spot groups with a well-defined maximum area were taken into account. We also analysed the covariance of the resulting model parameters and their correlations with the physical parameters of the sunspots and the ongoing solar cycle.}
   {Our results show that the temporal evolution of well-observed sunspot groups that reach at least 50~millionths of a solar hemisphere (MSH) at their maximum can be fitted surprisingly well with our model. Furthermore, we show significant asymmetry -- described by a skew parameter of fitted curves -- between the growing and decaying phases of analysed sunspot groups. In addition, we found a weak correlation between the values of skew parameters and the maximum area of sunspot groups and their hemispherical latitude.}
   {}


   \maketitle
%
\section{Introduction}\label{sec:intro}

Sunspots and the variation of these magnificent magnetic features on the solar surface are one of the key indicators of strength for solar activity, as sunspots are locations of large-scale magnetic field emergence. Many key characteristics of sunspots can be listed as important for analysis, such as their number, their location, and their spatial evolution. Sunspots and sunspot groups show very little stability, they evolve continuously and their evolution is a good indication of the changes in the magnetic field, the key physical quantity describing and driving the dynamical and thermodynamical state of the solar atmosphere and solar wind.

A popular and useful way to characterise the changes of sunspots and sunspot groups is their temporal profile (the variation of their daily area), as this is tightly connected to the variation of the magnetic flux, hence the time-dependent changes in the area reflect the development of the magnetic field of the Sun. For a comprehensive review of earlier work related to the analysis of spot development and decay, and the characteristics of spot groups, see, for example  \citet{2015LRSP...12....1V}.

Nowadays, it is widely recognised and accepted that the large-scale magnetic structure of sunspots is formed at the base of the convection zone \citep{1980Natur.287..616S}. At the same time, alternative theories have been suggested to explain the appearance of large-scale magnetic structures due to the coalescence of the small-scale turbulent magnetic field in the near-surface shear layer \citep{2005ApJ...625..539B}. In general, the accepted idea is that the emergence of magnetic flux is driven by buoyancy, while the decay results from the impact of turbulent erosion with the environment  \citep{1997ApJ...485..398P}.

Since the large-scale flux is generated by the solar dynamo, the observed characteristics of flux emergence and that of the subsequent decay provide vital clues as well as boundary conditions for dynamo models. Throughout their evolution, active regions (AR) are centres of magnetic activity, with the level and type of activity phenomena being dependent on the evolutionary stage of the AR. The area and position data of sunspots are widely used to analyse various aspects of solar activity, such as emergence of the magnetic field, growth, and decay of sunspots as well as the connection between the structural development of a sunspot group and its flaring capability. In this study we focus on the evolution of dominantly bipolar ARs, that is sunspot groups. The aim of our investigation is to develop a novel, reliable, and statistically sound technique to model the temporal variations of low resolution sunspot groups, which will be a foundation of future analysis dedicated to the understanding of the evolution of flux emergence and sunspot decay.

The analysis of the generation, evolution, and decay of ARs is one of the oldest and most studied areas of solar physics. While initially research on this topic focused mainly on the visible detection of the number of active regions and their large-scale evolution, the science of active regions received a different connotation once the role of ARs in the evolution of solar activity, atmospheric heating, and even space weather were established. It became clear that in order to understand the multi-faceted solar activity at various scales, one needs to understand the evolution of sunspots, their generation, evolution, and decay. In this respect, several studies addressed the problem of the formation and decay of active regions. Using digitised Mount Wilson data, covering the period from 1917-1985 \cite{1992SoPh..137...51H} analysed the daily growth and decay rates of sunspot group umbral areas. These rates were obtained to be distributed roughly symmetrically about a median rate of decay of a few $\mu$hemisphere~day$^{-1}$ ($\mu$Hem~day$^{-1}$ or MSH~day$^{-1}$). Percentage area change rates are reported to average 502\%~day$^{-1}$ for growing groups and -45\% day$^{-1}$ for decaying groups. These values are significantly higher than the comparable rates for plage magnetic fields because spot groups have shorter lifetimes than the plage regions. The distribution of percentage decay rates also differed from that of plage magnetic fields.

Using daily resolution data collected over 130 years by the Royal Greenwich Observatory (RGO) and the United State Air Force (USAF), \cite{2008SoPh..250..269H} found that the decay rate increases linearly for groups with areas increasing from 35~$\mu$Hem to 1000~$\mu$Hem. The decay rate they determined tends to level off for groups with areas larger than 1000 $\mu$Hem. This behaviour is very similar to the increase in the number of sunspots per group as the area of the group increases. Calculating the decay rate per individual sunspot gives a value of about 3.65~$\mu$Hem~day$^{-1}$, with little dependence upon the area of the group. These authors determined that the above values support the theory that predicts that sunspot decay is driven by diffusion, with a diffusion coefficient of approximately 10~km$^2$~s$^{-1}$. The same authors also found that high latitude spots tend to decay faster than low latitude spots. 

The minimum and the maximum values of the annual mean growth rates of spot groups were studied by \cite{2011SoPh..270..463J} and their results show that these are $\sim$52\%~day$^{-1}$ and $\sim$183\%~day$^{-1}$, respectively, whereas the corresponding values of the annual mean decay rates are $\sim$21\%~day$^{-1}$ and $\sim$44\%~day$^{-1}$, respectively. The average value (over the period 1874-2009) of the daily growth rate is about 70\% more than that of the decay rate. The growth and the decay rates vary by about 35\% and 13\%, respectively, on a 60-year time scale.

The relationship between the maximum area and the growth and decay times was also investigated by \cite{2014SoPh..289...91G} through the analysis of the time variation of the area of a group. These authors found that, when looking at the variations of maximum area as a function of the solar cycle phase, the growth was obtained to be more chaotic than the decay. The decay process showed a typical initial behaviour of rapid decline of the area with time, before finally experiencing a smoother decay until disappearance. 

Finally, we should mention that all the previous studies listed here (except \cite{1992SoPh..137...51H} and  \cite{2014SoPh..289...91G})  investigated the evolution of individual sunspots only, not sunspot groups. However, since sunspots appear in groups, here address this shortcoming. 

The paper is organised as follows: in Section~\ref{subsec:dataselection}  we introduce the data sets used for our statistical investigation, the selection criteria, together with the model employed for analysing the fitting curves. Section~\ref{subsec:fittingformula}--Section~\ref{subsec:priors} is devoted to the analysis of the temporal changes of areas by means of Bayesian models. Our results are discussed in detail in Section~\ref{sec:results}. Finally, in  Section~\ref{sec:conclusion} we summarise our key findings.

\section{Data sets and model}\label{sec:data} 

\subsection{Data sets and sample selection}\label{subsec:dataselection}

\begin{figure}
  \centering
  \includegraphics[width=0.99\linewidth]{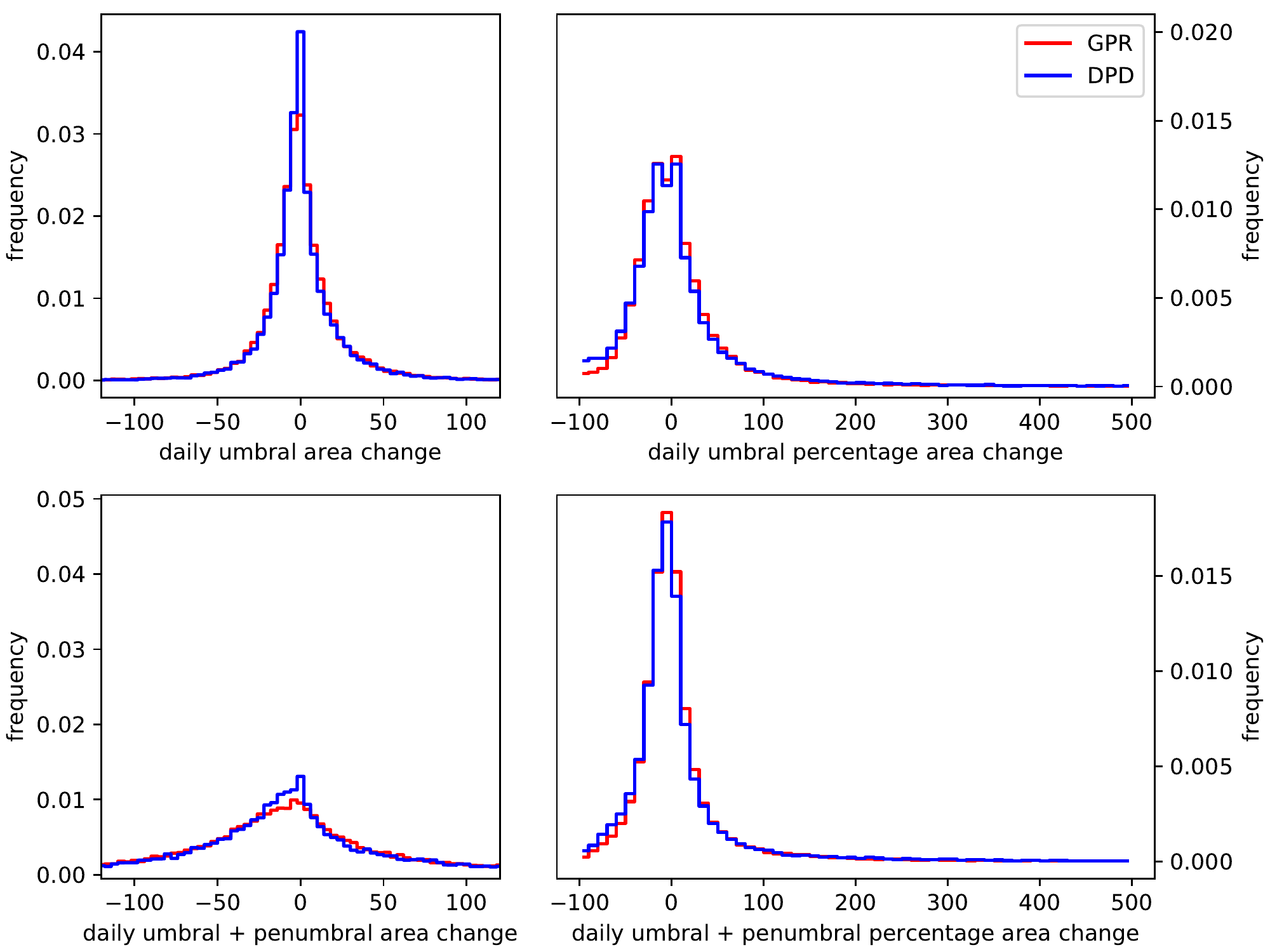}
  \caption{Histogram of daily sunspot group area changes (left column) and day to day percentage changes (right column) for the umbrae alone (top row) and umbrae + penumbrae (bottom row).}
  \label{fig:dailyarea}
\end{figure}

\begin{figure*}
\centering
\includegraphics[width=0.33\textwidth]{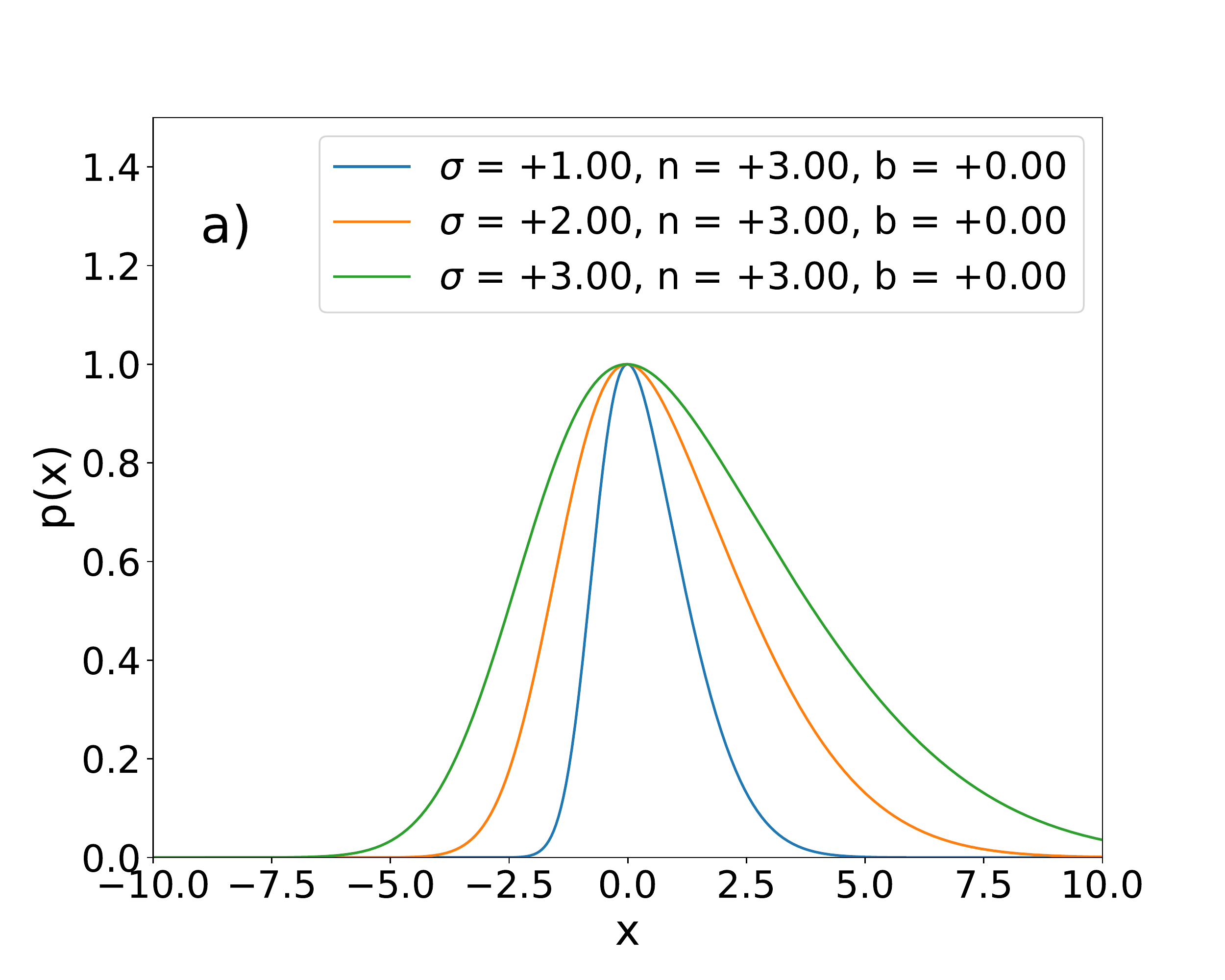}
\includegraphics[width=0.33\textwidth]{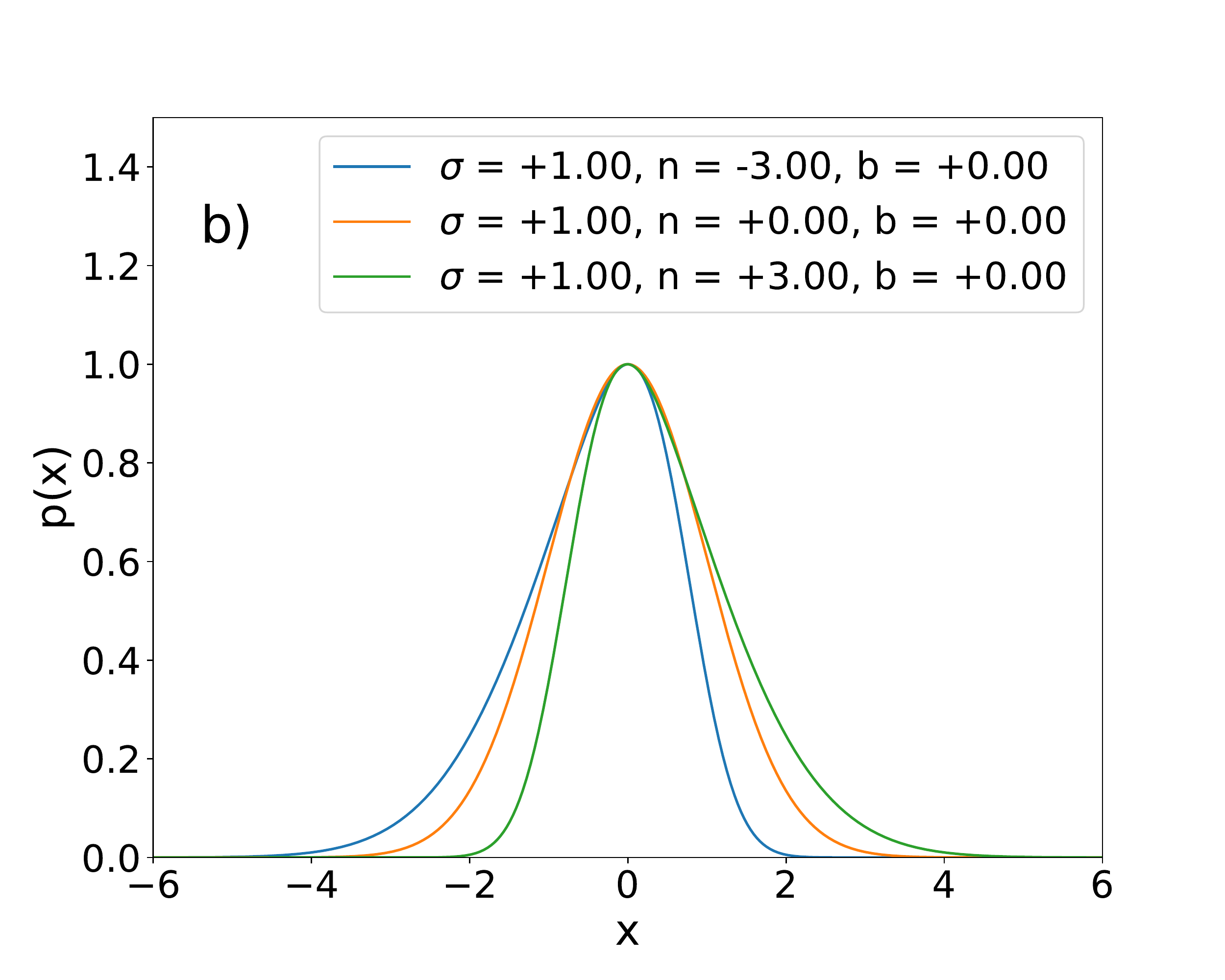}
\includegraphics[width=0.33\textwidth]{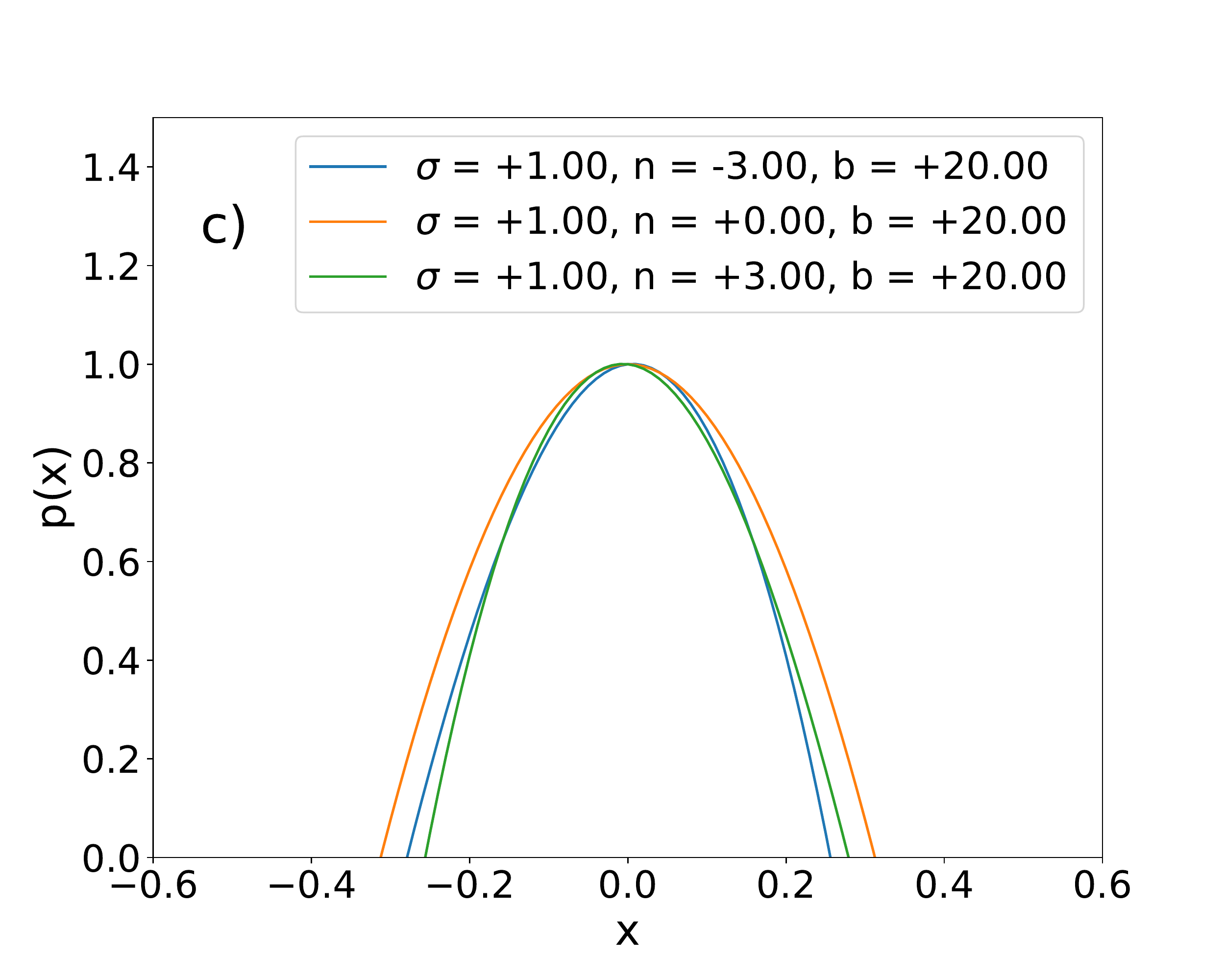}
\caption{Behaviour of the skew-normal-like model for different values of its parameters. \textbf{a)} the scale parameter, $\sigma$, has a very similar effect on the width of the curve as the variance of a Gaussian function. \textbf{b)} For positive (negative) values of the shape parameter, $n$, the curve is skewed to the right (left), i.e. the model rises more (less) rapidly than it decays. \textbf{c)} In case of $b > A$, the model is positive on a final interval only and its skewness is not as prominently visible as for $0 \le b \ll A$. For all three cases presented here we consider $a=1$ and $m=0$.}
\label{fig:model}
\end{figure*}

Our analysis is based on two well-established data sets. First, the Greenwich Photoheliographic Results (GPR) include daily measurements of sunspot positions and the areas of sunspots and faculae observed from 1874 to 1976 on a daily basis, covering 9 solar cycles. The Debrecen Photoheliographic Data (DPD) is a formal continuation of GPR, with 3 overlapping years containing daily measurements of sunspot positions and areas from 1974 to 2018  \citep{2016SoPh..291.3081B,2017MNRAS.465.1259G}. The measurements were made by white-light full-disk photographic observations and the archive consists of more than 150,000 digitised plates. The DPD dataset contains the area and position of each spot, the total area and the mean position of sunspot groups and the daily total area of all groups. In addition, the DPD dataset contains information about the penumbra region of each sunspot, where projection effects have been corrected. For both GPR and DPD, we use the revised data sets by \cite{2016SoPh..291.3081B} available online\footnote{\url{http://fenyi.solarobs.csfk.mta.hu/en/databases/DPD/}}. 

Several selection criteria were applied in connection to the chosen databases. First of all, we neglected all small spot groups where the total umbral and penumbral area at maximum did not reach 50~MSH (millionths of the solar hemisphere). Furthermore, for a better statistical relevance, we only included those groups that were observed at least 10~times, but not necessarily on 10 consecutive days. Next, we discarded those groups that had less than three records before 
and after the epoch of maximum area. Next, we excluded those sunspot groups for which the observations before and after the maximum area showed a small number of spots with nearly identical area as the maximum value because in this case the fitted curve to the growth or decay phase becomes flat, meaning that its intersection with the horizontal axis is unrealistically distant. Finally, we considered only those spot groups that had a single, well-defined maximal area, and filtered out groups with multiple strong local extrema, which might mean unusually turbulent evolution. We also note that due to the solar rotation (or other observational difficulties) there are groups whose emergence or disappearance is not visible, however, these groups were included in our database provided they satisfy our selection criteria. After applying all the above selection criteria, our sample database comprises 2215 GPR and 1645 DPD sunspot groups, where projection effects for sunspots area were corrected.

Throughout this study, if not otherwise noted, we use the term spot area for the combined umbral {\it and} penumbral area. Fitting the combined area is significantly easier as it shows much smoother variations than that of umbrae \citep{2014SoPh..289..563M}.

As a simple consistency check with the study by \cite{1992SoPh..137...51H}, we plot the histogram of daily umbral, as well as umbral + penumbral area and percentage changes in Fig.~\ref{fig:dailyarea}. Unsurprisingly, the distribution of daily umbral + penumbral area is significantly broader that those of the umbrae alone. On the other hand, percentage area changes tend to be smaller for the combined area, that is, it changes more smoothly over time.

\begin{figure*}[]
    \begin{tabular}{ccc}
        \includegraphics[width=0.3\linewidth]{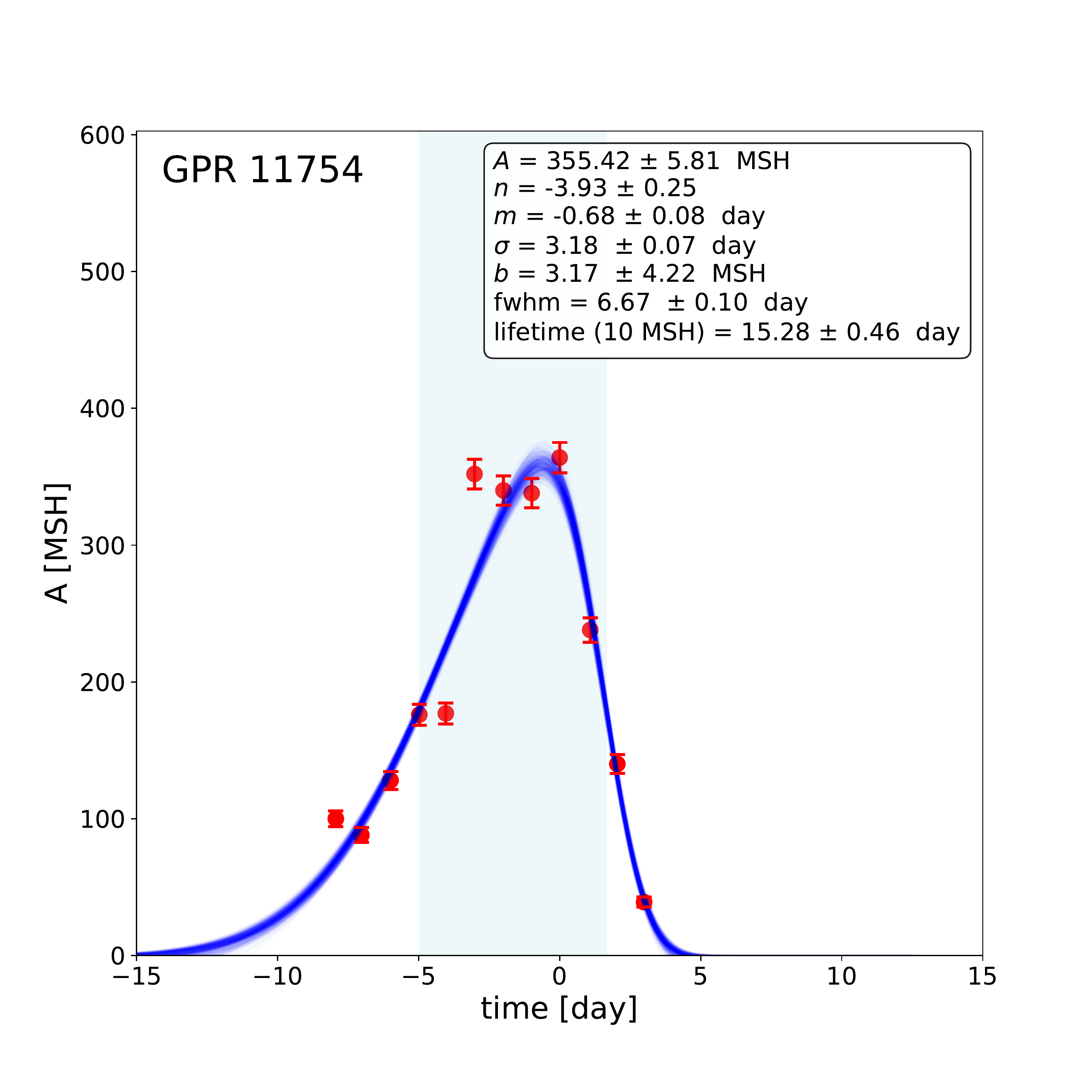}
        &
        \includegraphics[width=0.3\linewidth]{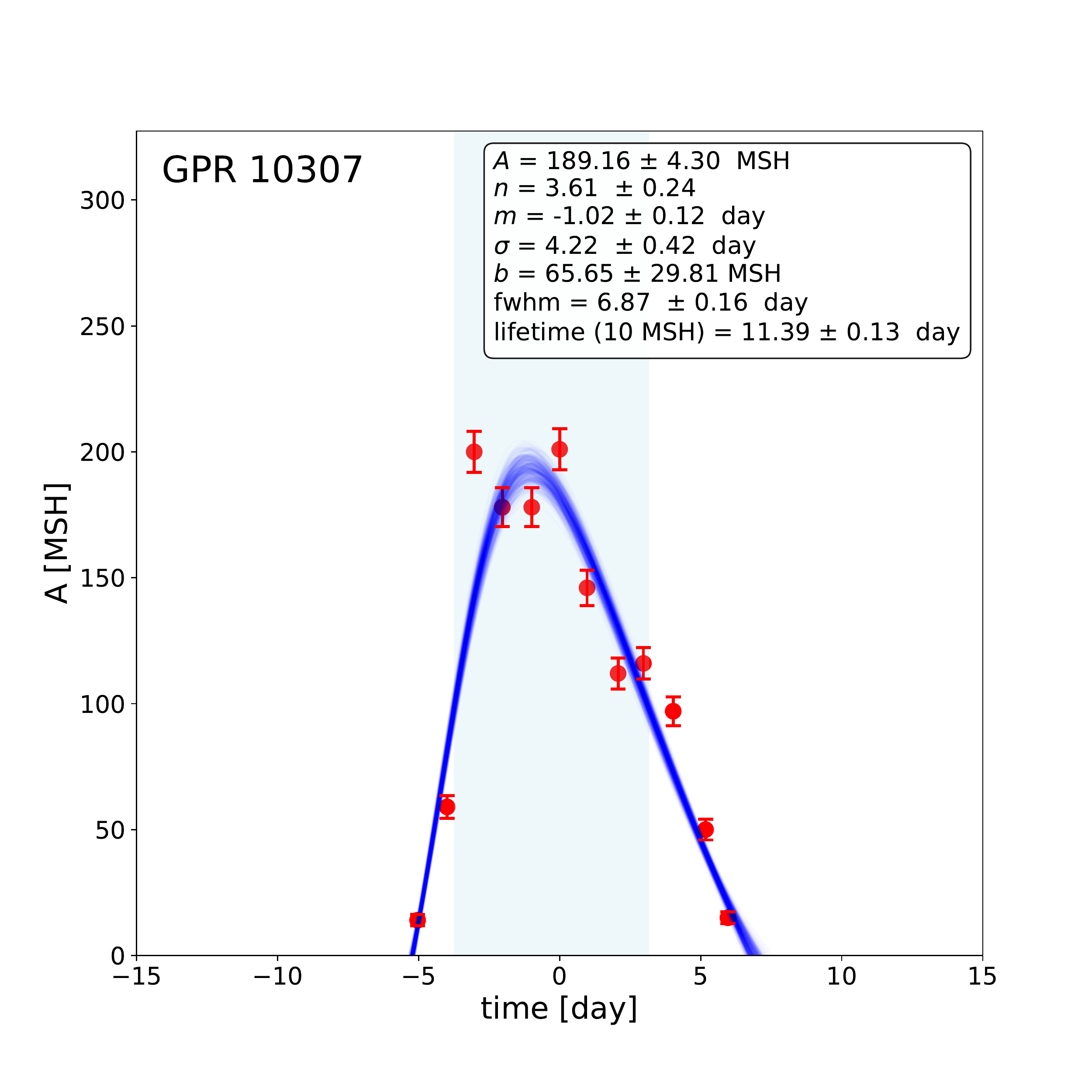}
        &
        \includegraphics[width=0.3\linewidth]{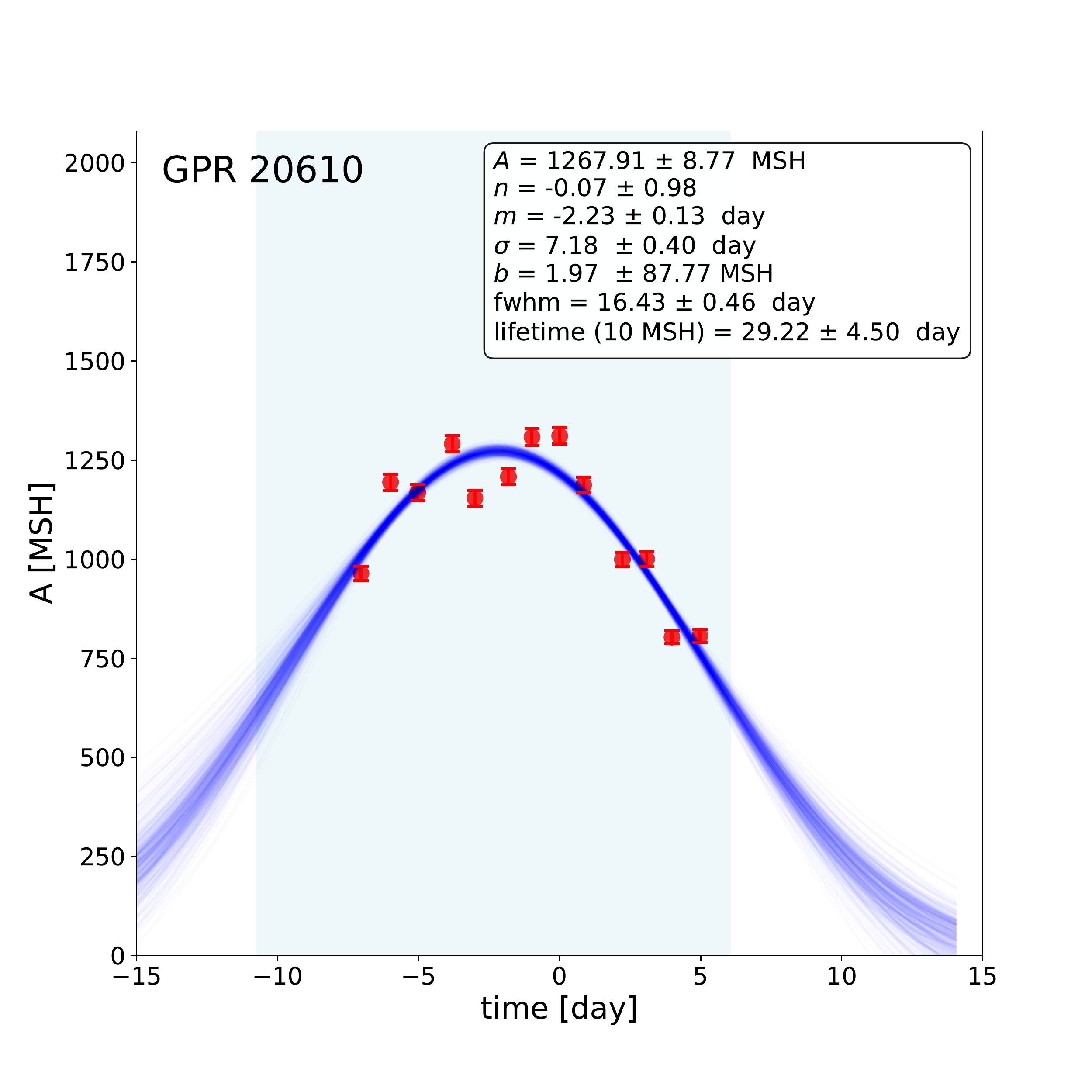}
        \\
        \includegraphics[width=0.3\linewidth]{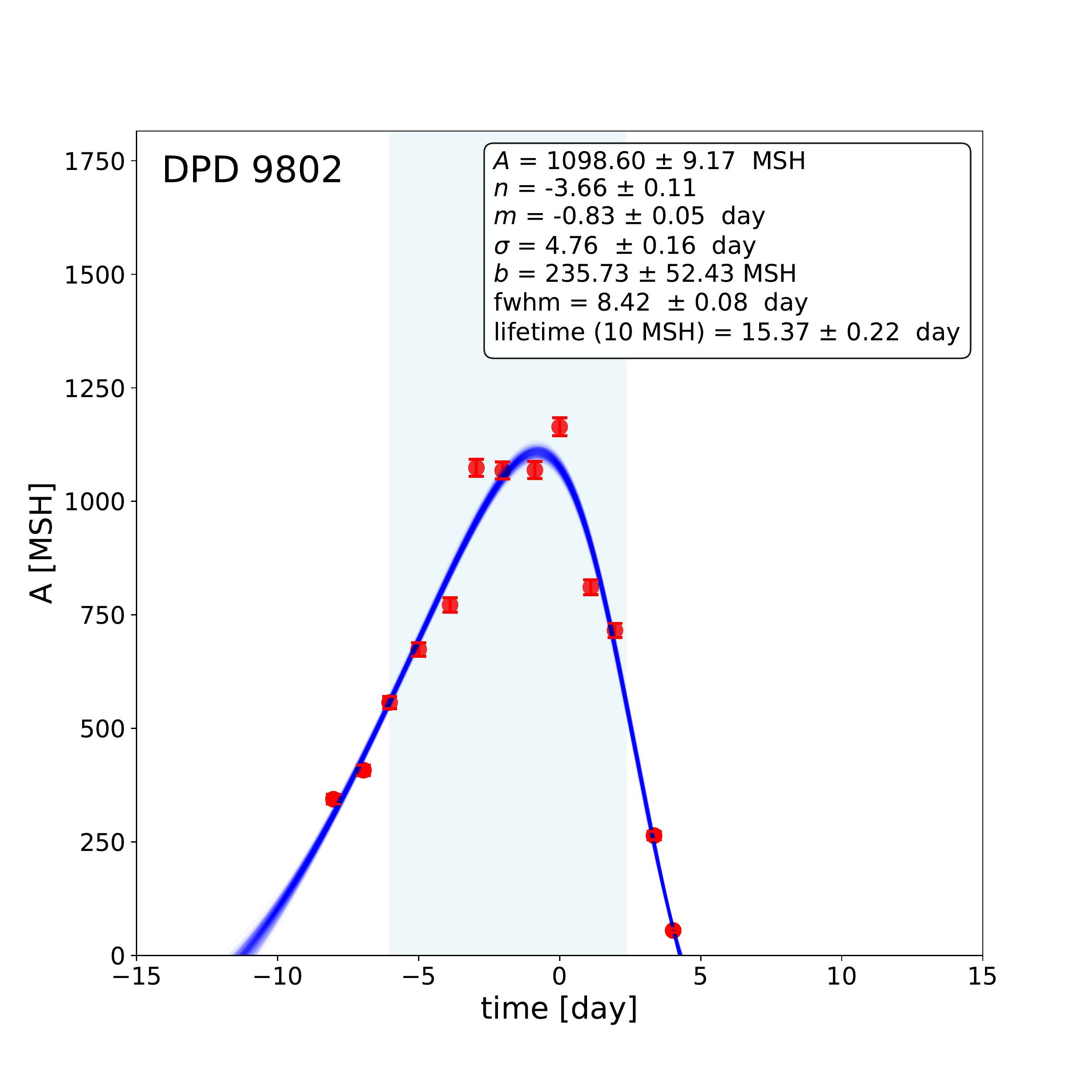}
        &
        \includegraphics[width=0.3\linewidth]{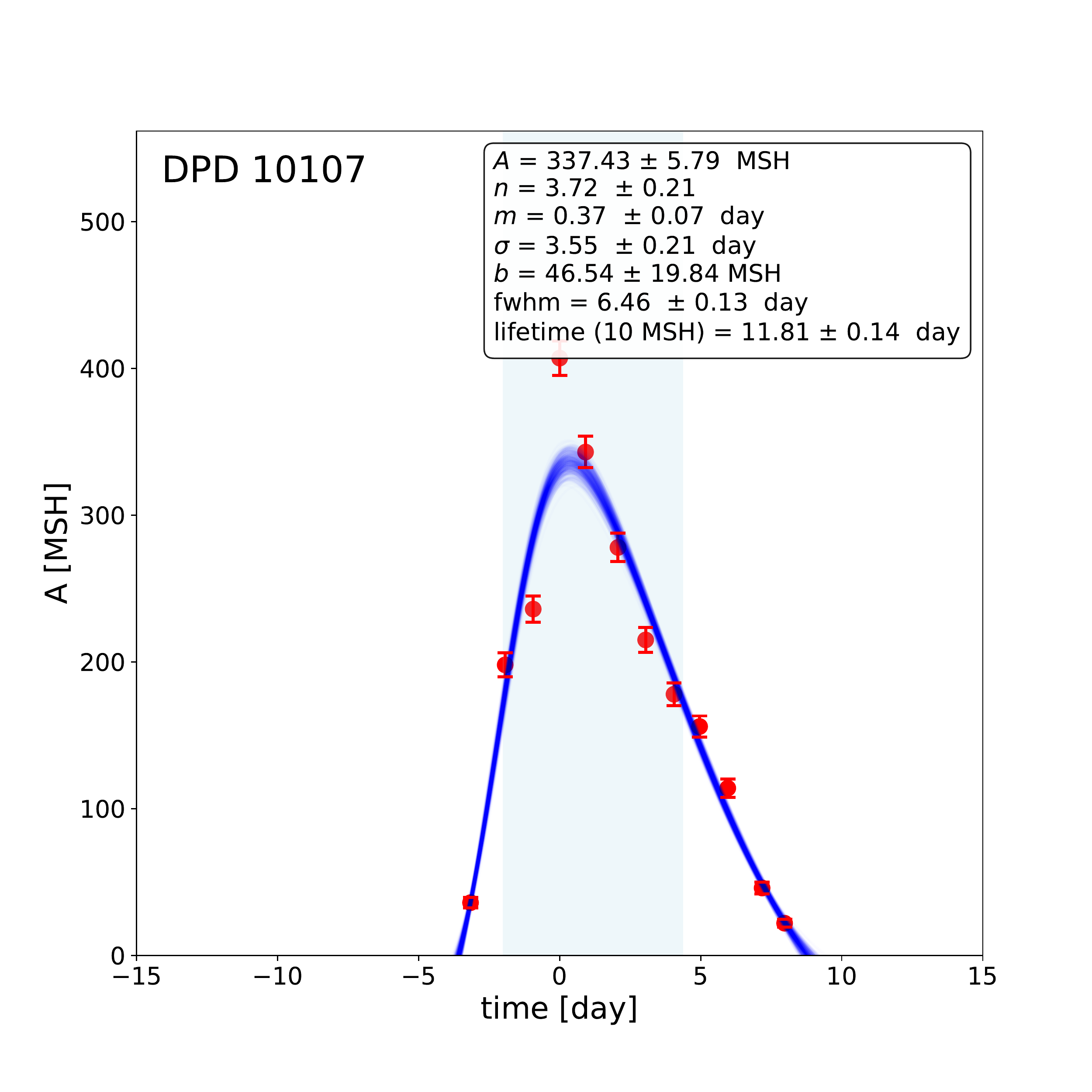}
        &
        \includegraphics[width=0.3\linewidth]{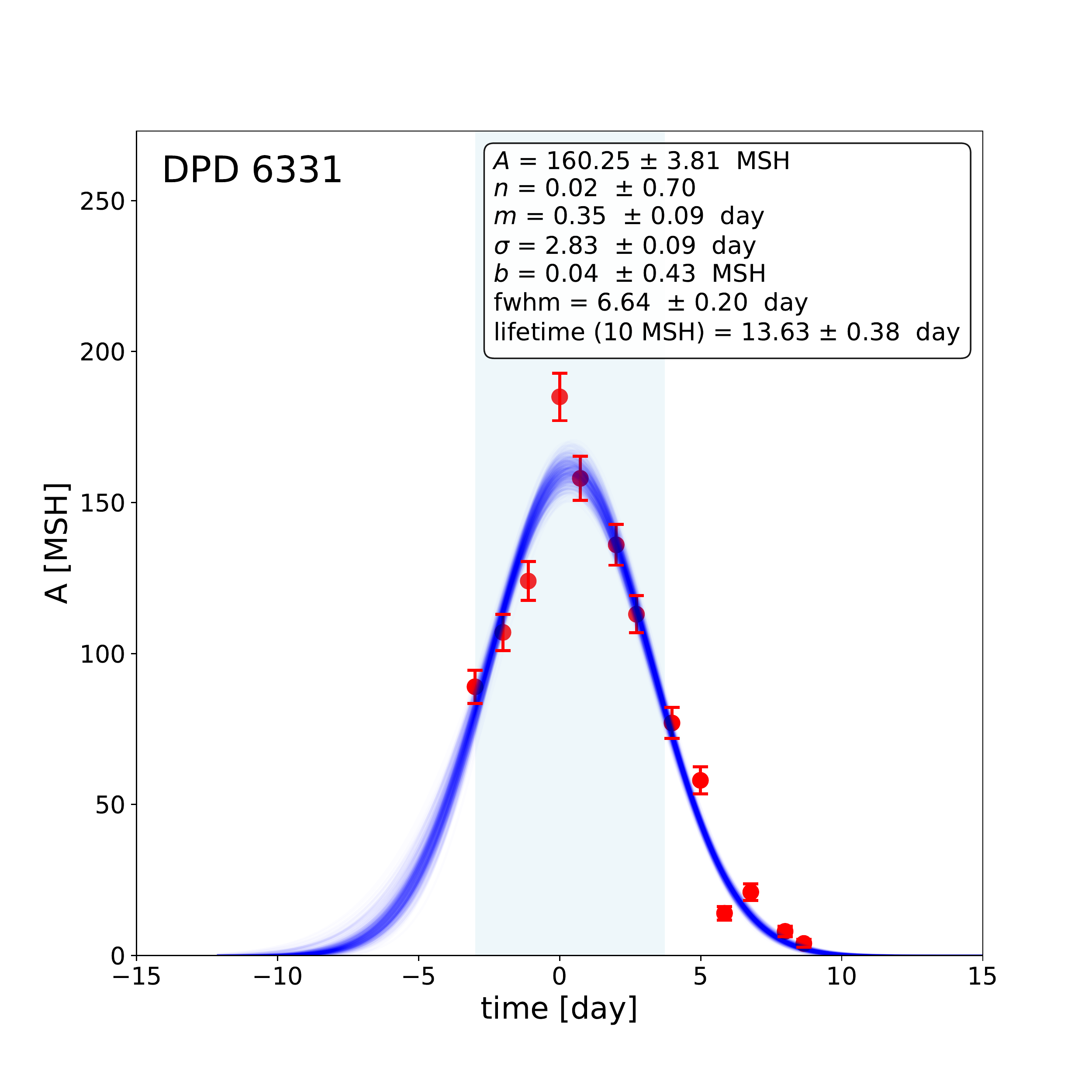}
    \end{tabular}
    \caption{GPR/DPD solar groups for negative/positive $n$ and a symmetric evolution ($n\approx 0$), as a representative example. The observed variations of the area with time are shown by red dots, while the fitting curves are represented by the blue lines for 500 MCMC realisations. The time on the horizontal axis is relative to the epoch of observed maximum. The error estimates of sunspots group area are calculated using Eq.~(\ref{eq:error}). The quantities displayed in the legend of the left-hand side of plots show the modes (most probable values) of the marginal posterior probability distributions of the model parameters and the associated errors. The background light-blue vertical strips mark the full width at half maximum (FWHM) calculated at the mode of all realisations.}
    \label{fig:gprdpdexamples}
\end{figure*} 

\subsection{A fitting formula for sunspot group areas}\label{subsec:fittingformula}

The primary goal of our study is to confirm or refute the hypothesis that the integrated area of sunspot groups is more likely to grow more rapidly over time during the phase of emergence than to fall more rapidly during decay and, that the more rapidly rising and more rapidly decaying groups form two distinct populations. Instead of looking at the time derivatives of the total sunspot group area, $A(t)$, we fit the values of the measured area as a function of time with a model modified from the skew-normal distribution \citep{1985Azzalini,1976OHagan}. To fit the temporal evolution of sunspot groups other formulae have been suggested by e.g. \citet{2011SoPh..273..231D, 2014SoPh..289..563M}. While very similar in shape, our formula has a weaker tail, c.f.~Fig.~7 of \citet{2014SoPh..289..563M}, which could be a better fit to rapidly growing and slowly decaying temporal profiles measured at higher temporal resolution with modern instruments. In addition, we allow for a constant term in the function because this enables us to fit temporal profiles where the time derivative of the area is non-zero at the time of appearance and disappearance of the spot group. While this formula yields better fits to low time resolution data, its validity on higher quality sunspot measurements is yet to be verified. Moreover, we engineered our formula to minimise the covariance of parameter estimates, especially for the scale and shape.

The skew-normal distribution is a generalised version of the normal distribution in the form of
\begin{equation}
    p(x) = A \cdot \exp\left({-\frac{(x-m)^2}{2\,\sigma^2}}\right) 
            \left[ 1 + \mathrm{erf}\left( \frac{n (x - m)}{\sigma \sqrt{2}} \right) \right] , 
\label{eq:skewnormal}
\end{equation}
where $m$ is a location parameter, which is different from the mean and the mode, $\sigma$ is a scale parameter that serves the same purpose as the $\sigma$ of the normal distribution, $n$ is a shape parameter that determines the skew towards negative ($n < 0$, rapidly rising) or positive ($0 < n$, rapidly decaying) values and $\mathrm{erf}(x)$ is the error function. The first three moments of the skew normal distribution are
\begin{eqnarray}
    \mu_1 &=& m + \sigma \delta \sqrt{2/\pi}, \label{eq:moments1}\\
    \mu_2 &=& \sigma^2 \left( 1 - \frac{2\delta^2}{\pi} \right),  \\
    \mu_3 &= &\frac{4-\pi}{2} \frac{(\delta \sqrt{2/\pi})^3}{(1 - 2\delta^2 / \pi)^{3/2}},\label{eq:moments3}
\end{eqnarray}
where the quantity $\delta$ is defined as
\begin{equation}
    \delta = \frac{n}{\sqrt{1+n^2}}.
\end{equation}
The value of the mode (in the case of $m=0$) of the skew-normal distribution cannot be calculated analytically but it is unique and a fitting formula is given by \cite{Azzalini2014TheSA} in the form of
\begin{equation}
    \tilde{m} = \delta \sqrt{\frac{2}{\pi}} - \frac{\mu_3}{2}\sqrt{1 - \frac{2 \delta^2 }{\pi}} 
              - \frac{\mathrm{sgn}(n)}{2}\exp\left( -\frac{2\pi}{\left| n \right|} \right).
\end{equation}
From Eqs.~(\ref{eq:moments1}-\ref{eq:moments3}) we see that the parameter $n$ not only determines the skewness of the distribution, but also affects other moments as well. As a result, when fitting Eq.~(\ref{eq:skewnormal}) to some data, one can expect significant covariance among the free parameters. To cancel out the covariance as much as possible, we defined our parametric fitting formula from the canonical skew-normal distribution to be
\begin{equation}
    q(t) = \frac{A + b}{\tilde A} \exp\left({-\frac{(t-m+\tilde m)^2}{2 \tilde \sigma^2}}\right) 
    \left[ 1 + \mathrm{erf}\left( \frac{n (t - m + \tilde m)}{\tilde \sigma \sqrt{2}} \right) \right] - b,
\label{eq:themodel}
\end{equation}
where $\tilde m$ is the mode of the original skew-normal distribution and $\tilde A = p(\tilde m)$ is the value of the skew-normal distribution at its mode. The parameter $b$ is a positive constant which can be used to adjust the baseline of the curve. This parameter turns out to be necessary to properly fit the low resolution data. The amplitude of the model is compensated for $b$, so that the maximum does not change when the baseline changes. We revisit the treatment of this constant in Sec.~\ref{subsec:priors}. Since $b > 0$, only the positive part of the model is considered. In Eq. (\ref{eq:themodel}) we substituted $x$ by $t$ to denote time and the modified scale parameter takes the form of
\begin{equation}
    \tilde \sigma ^2 = \frac{\sigma ^2}{1 - 2\delta^2/\pi}.
\end{equation}

The fitting formula defined by Eq. (\ref{eq:themodel}) is plotted in Fig.~\ref{fig:model} for a few interesting cases. When $b = 0$ the curves never intersect the horizontal axis and they rise and fall smoothly towards the wings. This case is a good model of spot groups which appear slowly and start to grow more rapidly afterwards. On the other hand, curves with $b > 0$, and particularly with $b > A$, as is illustrated in Panel~c of Fig.~\ref{fig:model}, are better fits to spot groups which appear very quickly and their area grows rapidly during the first few days, but the growths slow down afterwards. The distinction between the two types of groups is purely empirical but necessary to properly fit the data. Higher quality and better time resolution data would help confirm or refute the $b > 0$, and especially the $b > A$ cases.

\subsection{Bayesian model fitting}\label{subsec:bayesian}

According to our experience, when attempting to fit Eq.~(\ref{eq:themodel}) to the temporal profiles of sunspot group area at low time resolution, commonly used maximum likelihood parameter estimation methods often diverge or fail catastrophically by converging to completely wrong results. In addition, asymptotic errors derived from the Hessian are often very misleading due to the non-quadratic minima of the $\chi^2$ surface. For this reason, we decided to solve the fitting problem by constructing a Bayesian model and estimate the joint posterior probability distribution of the parameters $A, m, \sigma$ and $n$ with Markov Chain Monte Carlo (MCMC) sampling. This approach not only can circumvent the iterative fitting problem, but also yields proper error estimates and covariances for all parameters. 

Bayesian models describe the joint probability distribution of the model parameter in the light of measured data. This can be expressed as the conditional probability
\begin{equation}
    P(\theta \given D) = \frac{P(D \given \theta) P(\theta)}{P(D)},
\label{eq:bayes}
\end{equation}
where $\theta = \left\{ A, m, \sigma, n \right\}$ represents the vector of model parameters and $D = \left\{ A_t \right\}$ denotes the measured value of sunspot group area at each epoch. The term $P(\theta \given D)$ in Eq.~(\ref{eq:bayes}) is the joint posterior probability distribution of the model parameters which tells us how much a certain choice of model parameters is supported by observations. In general, this probability distribution cannot be calculated analytically but the Markov Chain Monte Carlo method can be used to generate samples from it. The resulting Markov chain will consists of random realizations of the $\theta_i$ vectors such that the parameter space will be sampled proportionally to $P(\theta \given D)$. The Markov chain sampled from the multivariate posterior distribution can be easily marginalised over all other parameters to obtain the distribution and the moments of any individual model parameter. In practice, this is carried out by analysing the histogram of the components of the $\theta_i$ vectors. In addition, one can easily determine the covariance of the model parameters by calculating the covariance matrix of the $\theta_i$ realizations. As an alternative to MCMC sampling from the posterior distribution, there are various gradient descent algorithms able to quickly determine the location of its maximum, that is the best fit parameters. The latter method is called maximum a posteriori estimation, however, this method does not yield uncertainties and it is less robust than a full MCMC sampling.

In  Eq.~(\ref{eq:bayes}) the quantity $P(D \given \theta)$ is the usual likelihood function, the probability of measuring data $D$, if we choose a certain $\theta$ parametrisation of the model. Assuming independent measurements of spot group areas with normally distributed errors, the likelihood takes the well-known form of $P(D \given \theta) \propto \exp({-\chi^2})$, where 
\begin{equation}
\chi^2 = \sum_t \left[ \frac{A_t - q(t, \theta)}{\sigma_{A,t}} \right]^2.
\label{eq:chi}
\end{equation}
In general, the Bayesian model would enable us to very naturally incorporate varying and/or asymmetric measurement errors of the total spot group area into the likelihood function.
    
The quantity $P(\theta)$ is the prior probability distribution of the model parameters which should reflect all our believes about the model and its parameters. The choice of a prior distribution can sometimes be challenging as no generic prescriptions are available. If no \textit{a priori} information is available, as in our case, so-called trivial or non-informative priors should be used, usually in the form of uniform distributions over certain intervals. For instance, one can define a uniform prior on parameter $A$ between $0$ and some reasonably high value to cover the case of extreme high area groups, as well. Alternatively, one can use a log-normal prior if the assumption is that extreme high area sunspot groups are rare. In any case, the choice of prior should not affect the posterior distribution of the model parameters significantly but the data itself should dominate the posterior distribution via the likelihood function.

Finally, $P(D)$ is the prior predictive distribution which tells us, based solely on the model, what the distribution of the observed data should be. In theory, this can be calculated by integrating the probability $P(D \given \theta)$ over all possible $\theta$ but the analytic calculation is usually not possible. In practice, the value of $P(D)$ is only the normalisation factor of the posterior distribution which is indifferent when doing MCMC sampling from $P(\theta \given D)$ because it cancels out when calculating the transition probabilities.

\subsection{The choice of priors}\label{subsec:priors}

We choose the trivial uniform prior for all model parameters, except for $b$, for which an exponential prior is used. The choice of a trivial prior does not need much justification, whereas the choice of the exponential prior for $b$ was made to reflect our assumption that the group area tends to grow only slowly at the beginning, i.e. the time derivative of $A(t)$ is close to zero, which is not true when $b > 0$. This assumption prefers models with $b \approx 0$ even when observations of a sunspot group are missing when the group is still or already too small to detect. On the other hand, in certain cases when the data strongly prefers it, large values of $b$ are sampled with high posterior probability. Unfortunately, no upper limits or exact zero values for group area are available in the data sets to fit the tails more reliably. 

As a result, our priors are given by
\begin{eqnarray}
    P(A) & = & \mathcal{U}(0, 10000), \\
    P(m) & = & \mathcal{U}(-30, 30), \\
    P(\sigma) & = & \mathcal{U}(0, 30), \\
    P(n) & = & \mathcal{U}(-20, 20), \\
    P(b) & = & \exp(-0.01 b).
\end{eqnarray}

\subsection{Uncertainty of sunspot group areas}\label{subsec:uncertainlyarea}

Neither the GPR nor the DPD data sets provide error estimates on sunspot group area measurements. When group area is determined algorithmically from images, the core of the problem is to reliably identify the boundary of the group, hence the absolute measurement error should be proportional to the circumference of the group instead of its area. Based on this consideration, we assumed normally distributed errors with a variance of
\begin{equation}
    \sigma_{A, t} = \sqrt{0.3 \, (A(t) + 1)},
    \label{eq:error}
\end{equation}
where the $+\,1$ is a softening term for very small values. This assumption of uncertainty will allow more variance when fitting the peak of $A(t)$, which is likely to be measured with larger absolute error and, at the same time, will constrain the growing and decaying tails better.


\begin{figure}[]
    \centering
    \includegraphics[width=0.8\linewidth]{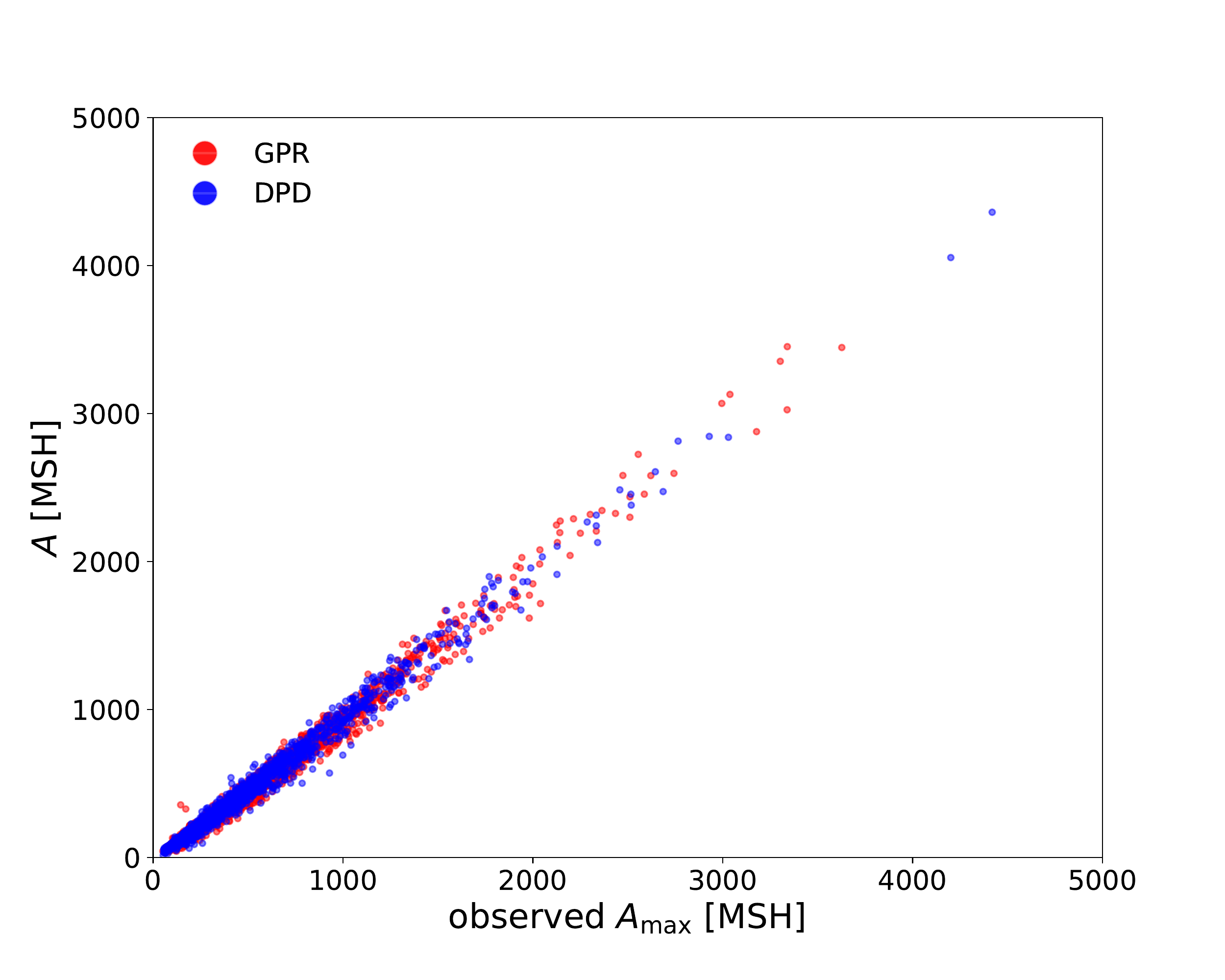}
    \caption{The most probable value of $A$ from 500~MC realization of the model plotted against the observed maximum spot group area, for every sunspot group of both data sets. The scatter plots for the GPR and DPD are consistent.}
    \label{fig:scatter-obsa-a}
\end{figure} 

\begin{figure}[]
    \centering
    \includegraphics[width=1.1\linewidth]{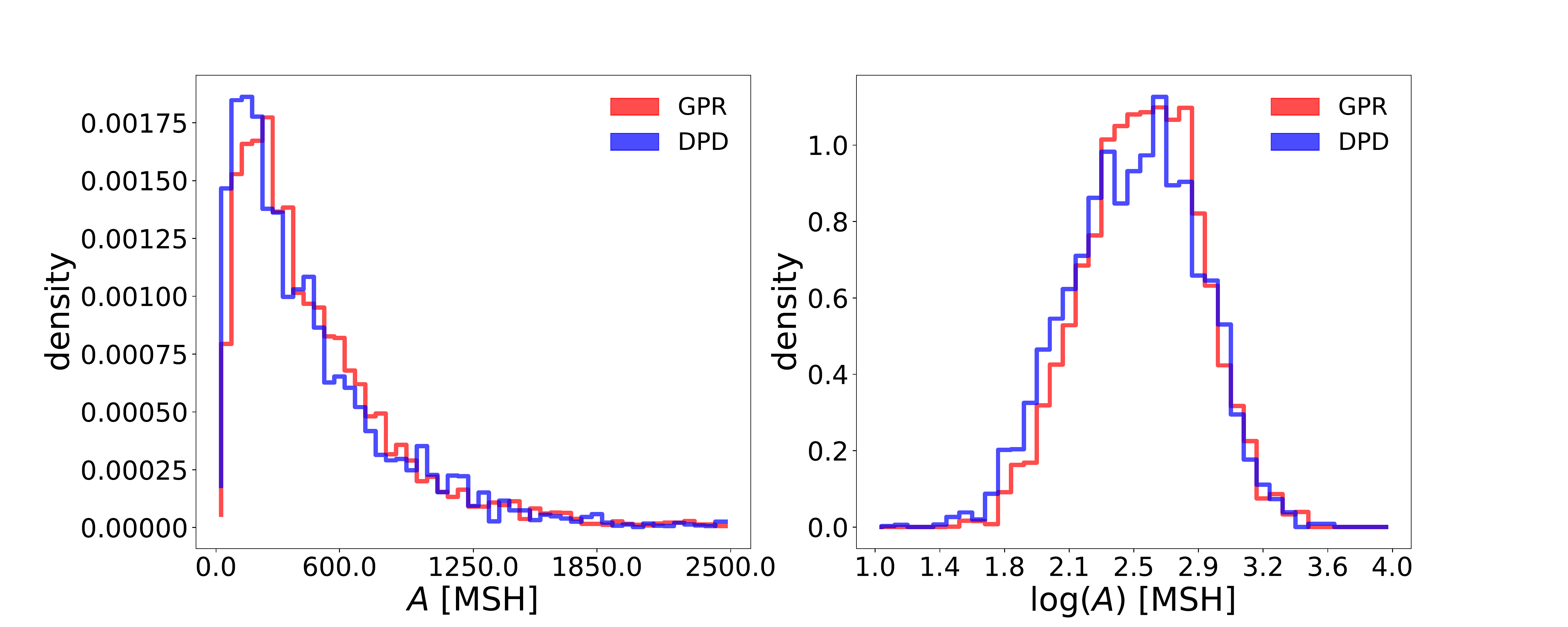}
    \caption{Normalised histograms of the parameter $A$ of the fitted model of all MC realizations for every sunspot groups of each data set. In the left-hand panel the histogram is shown in linear scale, while on the right we plot the distribution of the logarithm of the  groups' area since this allows better visualisation of small and large values. The histograms for the GPR and DPD are in general consistent, however small differences are visible for particular ranges of sunspots' areas.}
    \label{fig:hist-a}
\end{figure} 

\section{Results}\label{sec:results}

The model presented in Section 2 was implemented by using the \texttt{pymc}~2.3 library  \citep{2015arXiv150708050S}, which employs an adaptive Monte Carlo method to sample the posterior joint probability distribution of the model parameters. For each sunspot group, $10^5$ realization were executed with a burn-in period of $5\times 10^4$. The sample was thinned by 100 to eliminate auto-correlations.

In what follows, we study the evolution of spot groups through histograms (one and two dimensional) and scatter plots. Histograms showing the distributions of certain parameters for the data samples, are computed from all MCMC realizations of all of the fitted parameters for every spot group, as opposed to taking the mean or the mode (maximum aposteriori) of the posterior probability density function. On the other hand, for scatter plots, where individual sunspot groups are represented, we take the mode of the MCMC realizations on a per spot group basis.

\subsection{Representative examples of MCMC model fits}\label{subsec:examplefit}

The results of the MC sampling for six typical sunspot groups are plotted in Fig.~\ref{fig:gprdpdexamples} corresponding to the cases of negative, positive, and symmetric skew parameter, $n$. Instead of just plotting the best fitting curve (i.e. the maximum \textit{a posterior} fit), we plot 500 different Monte Carlo realizations to illustrate that the data constrain the model very well and the model itself provides an accurate description of the changes of sunspots' area.

Although we do not plot the covariance of parameters, the parameters are better constrained in the case of high values of $n$ -- when the fitting curves resulting from the MCMC realisations are very close to each other -- than in the case of symmetric or almost symmetric groups that correspond to $n\approx 0$, where the fitting curves are more dispersed. This covariance is purely due to data, as we used a model which was formulated specifically to eliminate parameter covariances due to their functional form.

As we mentioned before, the databases did not provide an error estimate on total group are (umbra + penumbra), nor on the individual spots. Our empirical estimation of the area error is based on the manual analysis of spot groups which could be fitted well and determined from the scatter around the best fit model. Furthermore, we assume a roughly circular geometry for the spots and expect the area measurement error to be proportional to the length of its contour. This suggests the square root in the formula. The error estimates of the fitted model parameters are determined from the variance of the Monte Carlo realizations and these errors are shown for each case listed in  Fig.~\ref{fig:gprdpdexamples}.

Once the fitting curves are obtained for our sample, we use the goodness of fit $\chi^2$ value defined by Eq. (\ref{eq:chi}) for the mode values of parameters. As a limiting threshold we consider only those datasets that have their $\chi^2$ value less than 50. This definition would help us deselect outlier spot groups that would arise from non-standard datasets due to, e.g. erroneous observations or other raw data processing circumstances. As a result, the number of samples for both datasets is reduced by approximately 10\%.

\subsection{Maximum area of sunspot groups}\label{subsec:maxarea}

For a consistency check of model fits, we plot the mode of $A$ over all MC realizations of each spot group as a function of the observed maximal group area in Fig. \ref{fig:scatter-obsa-a}. The scatter plot of $A$ versus $A_\mathrm{max}$ suggests that the model fits are highly consistent with the measurements of maximal area of spot groups.

According to the plot shown in Fig. \ref{fig:hist-a}, the maximum total area of the umbra and penumbra of groups follows a distribution that resembles a log-normal distribution. In order to better visualise the small changes in the distribution of the area of sunspot groups, we plot the logarithmic variation of the area, as suggested by earlier studies \citep[e.g.][]{1997SoPh..176..249P,2015ApJ...804...68M,2019ApJ...871..187N}. As we specified earlier, in our analysis we neglected all small groups where the total (umbral and penumbral) area at its maximum did not reach 50~MSH. The histograms for the GPR and DPD are in general consistent, however small differences are visible for particular ranges of sunspots' areas; comparing the two figures it is clear the maximum is at approximately 300~MSH. Accordingly, the DPD observations give a larger number for smaller spots between $50 ~\mathrm{MSH} < A < 200~ \mathrm{MSH}$. For intermediate size groups, with areas between $300 ~\mathrm{MSH} < A < 1000 ~\mathrm{MSH}$, the GPR observations give a larger number, while for very large group areas the two databases give similar density, although in certain bins one can see an area excess in the DPD data. 

The peak of the distribution seen in Fig. \ref{fig:hist-a} corresponds to the second peak of the distribution of the logarithm of sunspots' area noted earlier by \citet{1968IAUS...35..174D} and subsequently confirmed by \citet{2015ApJ...804...68M, 2016ApJ...833...94N,2021ApJ...906...27N}. The reason for having only one peak in our data is due to the cut-off values in sunspot group area (50 MSH).


\subsection{Lifetime of sunspot groups}\label{subsec:lifetime}

\begin{figure}[]
    \centering
    \includegraphics[width=1.1\linewidth]{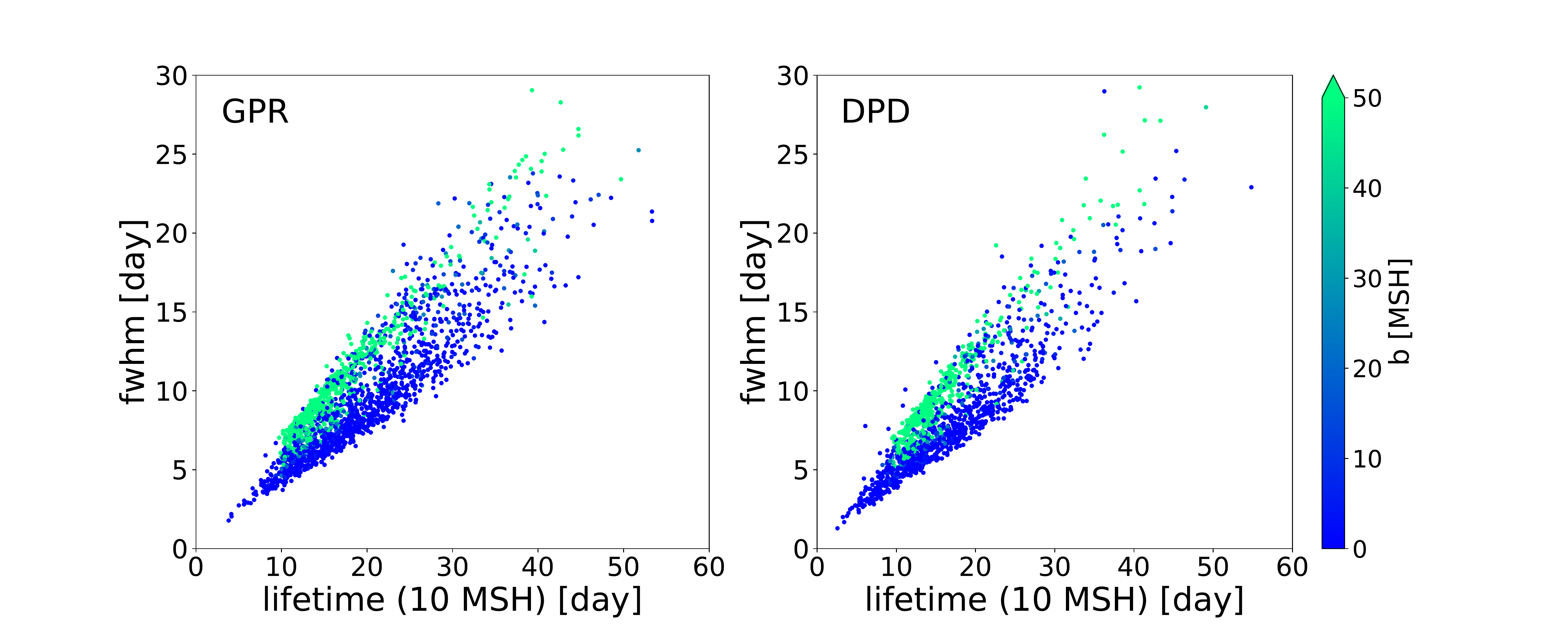}
    \caption{The correlation diagram of the FWHM and the threshold-based lifetime of sunspot groups.}
    \label{fig:colored-scatter-lifetime-fwhm-b}
\end{figure}

\begin{figure}[]
    \centering
    \includegraphics[width=1.1\linewidth]{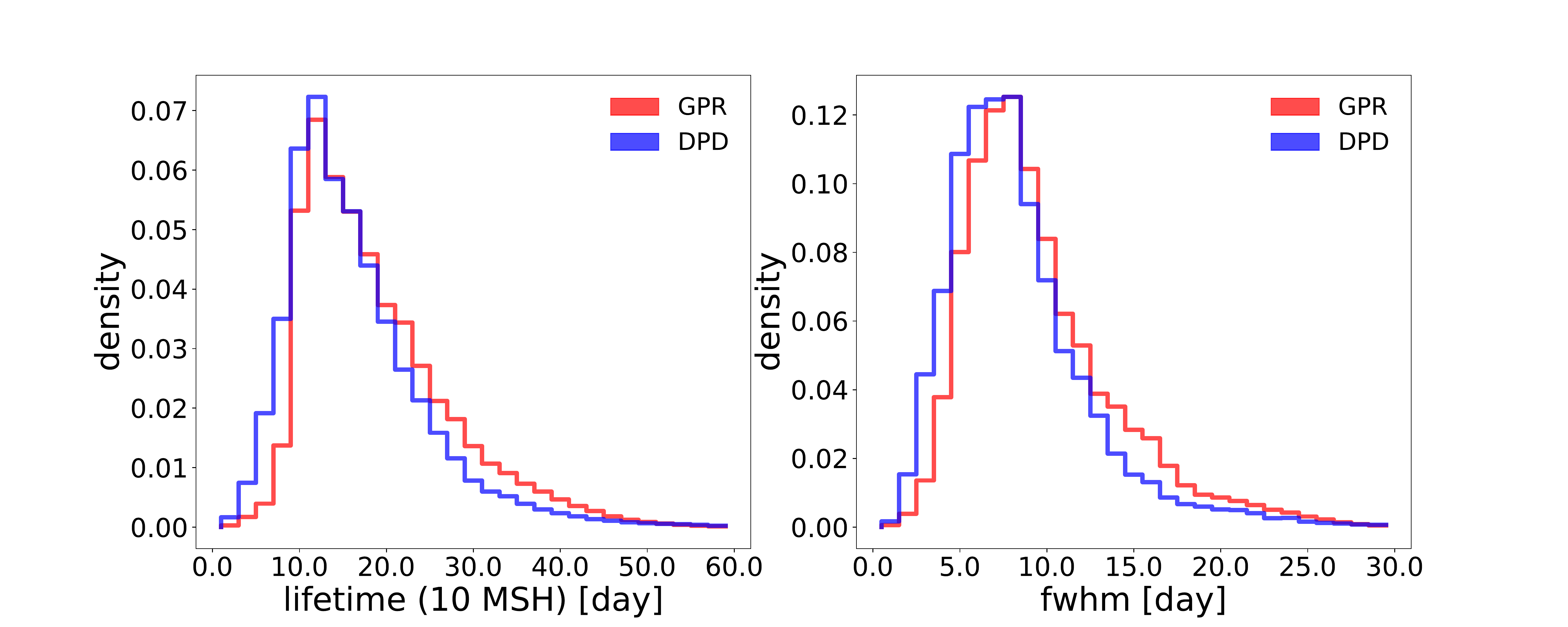}
    \caption{Normalized histograms of the threshold-based lifetime ({\it left-side}) and the FWHM ({\it right-side}) of the fitted model of all MC realizations for every sunspot groups of each data set. The histograms for the GPR and DPD datasets are in general consistent, however, small differences are visible for particular ranges of sunspots lifetime.}
    \label{fig:hist-lifetime}
\end{figure} 

\begin{figure}[]
    \centering
    \includegraphics[width=1.1\linewidth]{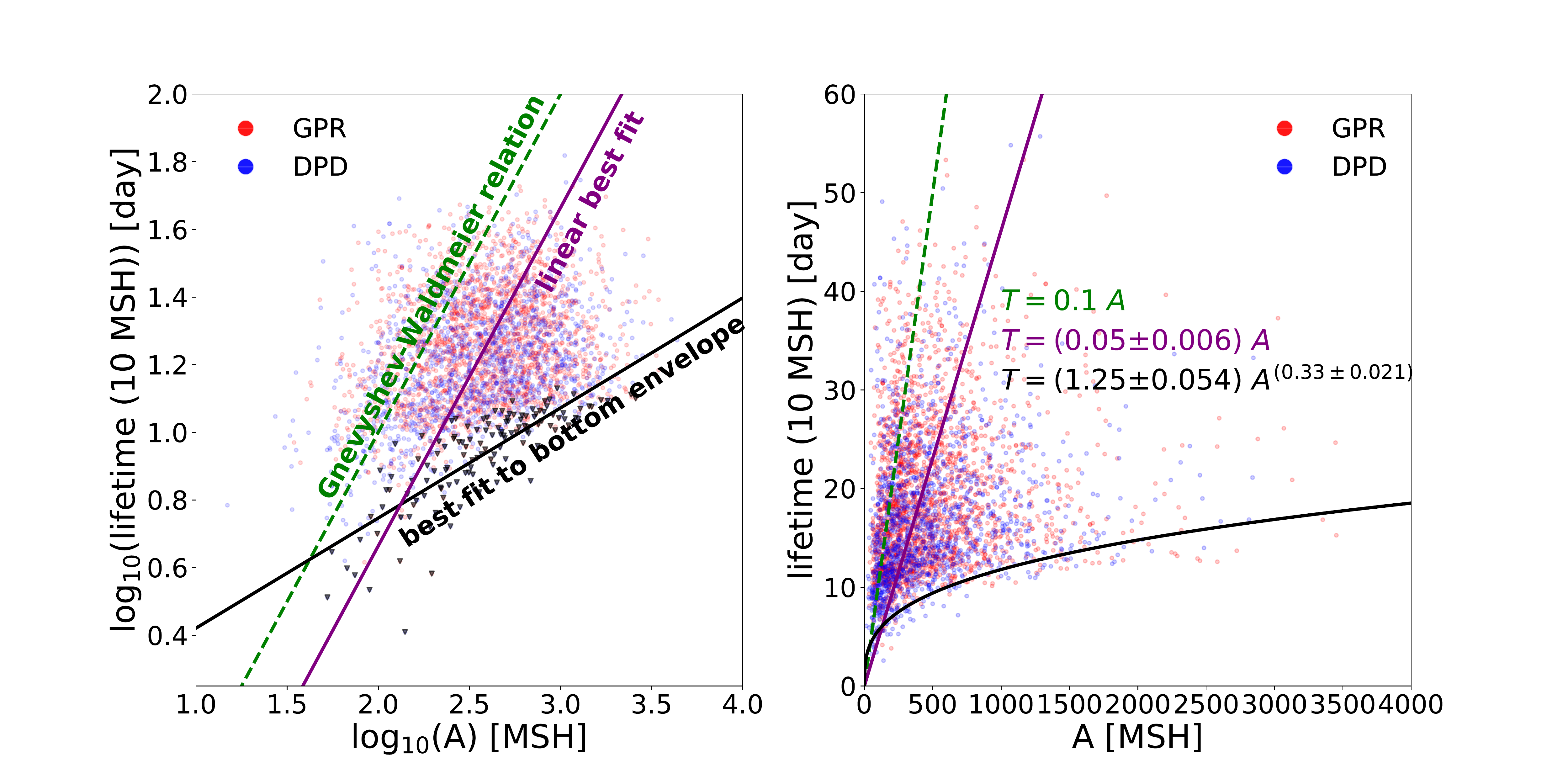}
    \caption{The lifetime of sunspot groups in terms of their area. In the left panel we plot the logarithmic values of the lifetime and the area, while in the right panel we plot their linear values. The dashed green line denotes the variation that corresponds to the Gnevyshev-Waldmeier rule, the purple solid line shows the best linear fit based on our data, while the black solid curve represents the lower envelop. The envelop curve has been obtained by fitting the groups shown by dark triangles. The errors shown in the equation of curves were derived from the standard deviation values. 
    }
    \label{fig:a-lt-fit}
\end{figure} 

\begin{figure}[]
    \centering
    \includegraphics[width=0.8\linewidth]{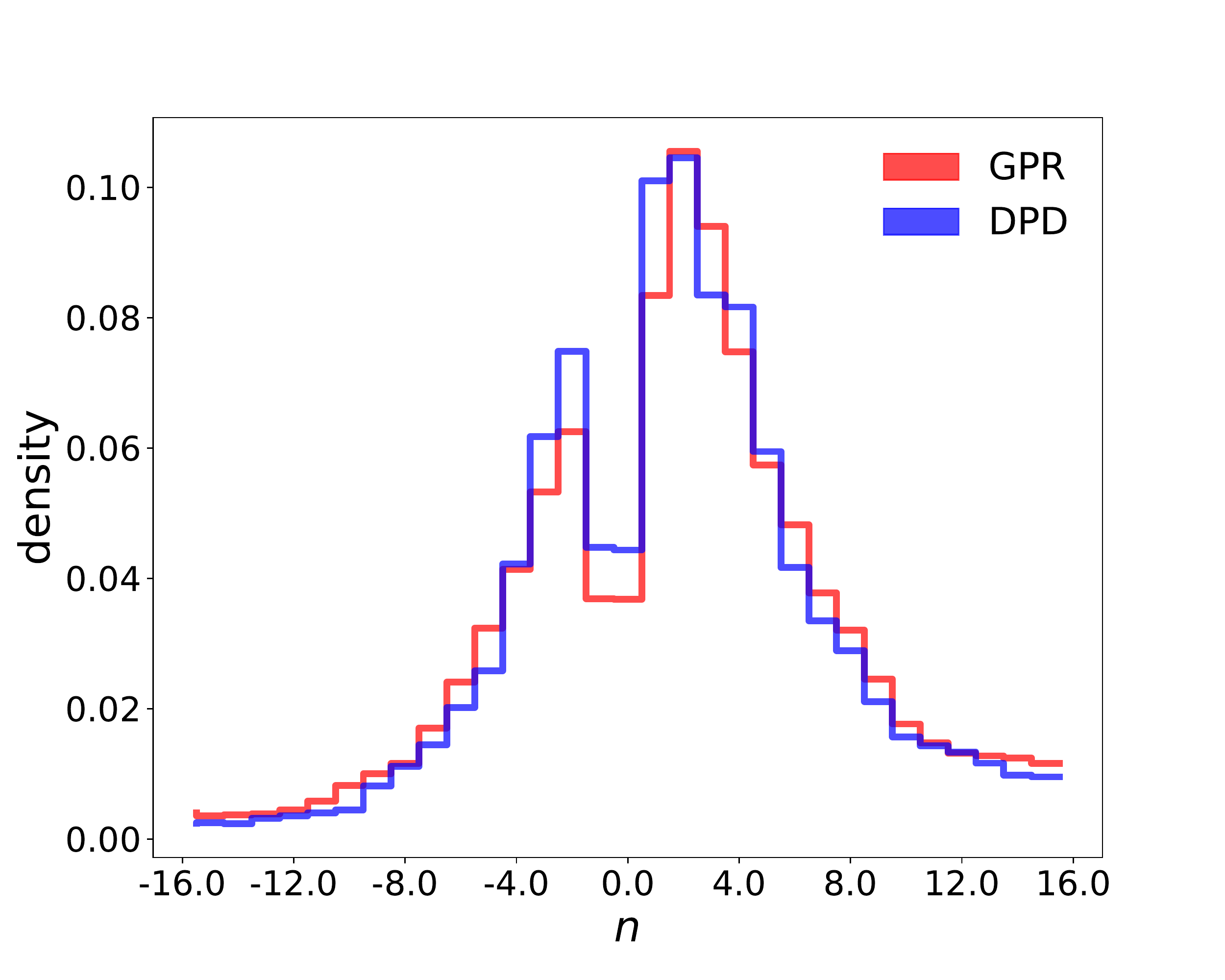}
    \caption{Normalized histograms of the skew parameter, $n$, of the fitted model of all realizations for every sunspot groups of each data set. The histograms for the GPR and DPD datasets are, in general, consistent, with small visible discrepancies.}
    \label{fig:hist-n}
\end{figure} 

There are several studies in the literature that discuss the possible relationship between the area of sunspots and their lifetime, one of the best known being the Gnevyshev-Waldmeier rule that states that the relationship between these two quantities is of the form $A_{max} = a \cdot LT$, where $A_{max}$ is the maximum sunspot group area during its lifetime in millionths of the solar hemisphere (MSH), $LT$ is the sunspot group lifetime in days, and $a$ is a constant parameter equal to $10$ MSH/day from the determined from the  Greenwich data for the period 1912–1934. However, in their investigation, these authors used groups whose growth and decay occurred during a single rotation in the field of view (half of the solar rotation), and the so-called recurrent groups observed during more than one solar rotation, i.e. groups with very small and very large lifetime. In these studies the exact observation of the appearance and disappearance of a sunspot group was a necessity. In contrast, here we determine the lifetime of groups by means of fitting curves and exact observation of the time of appearance /disappearance was not necessary. In this way we can study not only sunspot groups with very short and/or very long lifetime, but also those of intermediate lifetime. The exact methodology for determining the lifetime will be described later.

The Gnevyshev-Waldmeier rule has been investigated by many authors \citep[see e.g.][]{1997SoPh..176..249P,2019AstL...45..396N}. On the other hand \citet{1993A&A...274..521M} found that the decay rate of the sunspot area varies according to a parabolic law leading to a dependence of the form $T \propto \sqrt{A}$. Observationally, the lifetimes of sunspot groups are roughly consistent with the Gnevyshev-Waldmeier rule, though a large scattering around the Gnevyshev-Waldmeier rule can be seen in Fig.~5 of  \citet{2010SoPh..262..299H}.

As we mentioned in Sec.~\ref{subsec:maxarea}, in literature there are two distinct sunspot group populations based on their area. This characteristic can also be recovered in their lifetime \citep{2016ApJ...833...94N}. Accordingly, the sunspot groups form two populations: small short-living groups (SSG) and large long-living groups (LLG). The populations are separated by life-time strictly (greater and less than 5 days; \citet{2016ApJ...833...94N}), the magnetic field of the largest sunspot (greater or less than 2000 G; \citet{2016ApJ...833...94N}), and rotation (single- and two-component fast and slow). The existence of two different populations of sunspot groups was also mentioned by \citet{2015ApJ...804...68M}.

To investigate the lifetime of sunspot groups we start from our fitted curves. We compare two possible methods to determine the lifetime of groups. In the first method we use the parameters of the fitted skew-normal Gauss profile. If $b=0$, the scale parameter of the model curve, $\sigma$, would be a good measure of characteristic life-time of a spot group. In the general case of $b \ge 0$, however, one can use the full-width at half-maximum (FWHM) of the curve as characteristic life-time. Instead of attempting to derive an analytic formula for the FWHM as a function of model parameters, it is much easier to determine this parameter numerically. As a second possibility, we define a reference area of $10~\mathrm{MSH}$ and measure the time interval between the two zero crossings of the fitted curve. To avoid non-realistic values due to the lack of information for the starting phase of groups' evolution  we choose to start our estimation from a $10~\mathrm{MSH}$ reference level because for cases corresponding to $b=0$, the tail of the Gaussian distribution crosses the horizontal axis at infinity, resulting in a spurious value of the lifetime. In this case the lifetime can be defined as the temporal length between the intersection of the distribution curve with the horizontal threshold level.
 
In Fig.~\ref{fig:colored-scatter-lifetime-fwhm-b} we plot the correlations of the FWHM and threshold-based lifetime for the two databases. In the ideal case the results of the two methods would be displayed along the straight line with the FWHM corresponding to approx. a half lifetime. However, in reality this correlation is far from being ideal. The difference could arise because groups do not evolve symmetrically, but it could also be because of the parameters {\it b} or $n$. For illustration, in Fig. \ref{fig:colored-scatter-lifetime-fwhm-b} the colour bar shows the values of the parameter {\it b}, with $b>50$ shown uniformly in green. The 'bi-modality' of data can also be observed in Fig. \ref{fig:colored-scatter-lifetime-fwhm-b}, that is due to the parameter $b$, this being more pronounced for larger values of $b$.

\begin{figure*}[]
    \centering
    \includegraphics[width=0.49\linewidth]{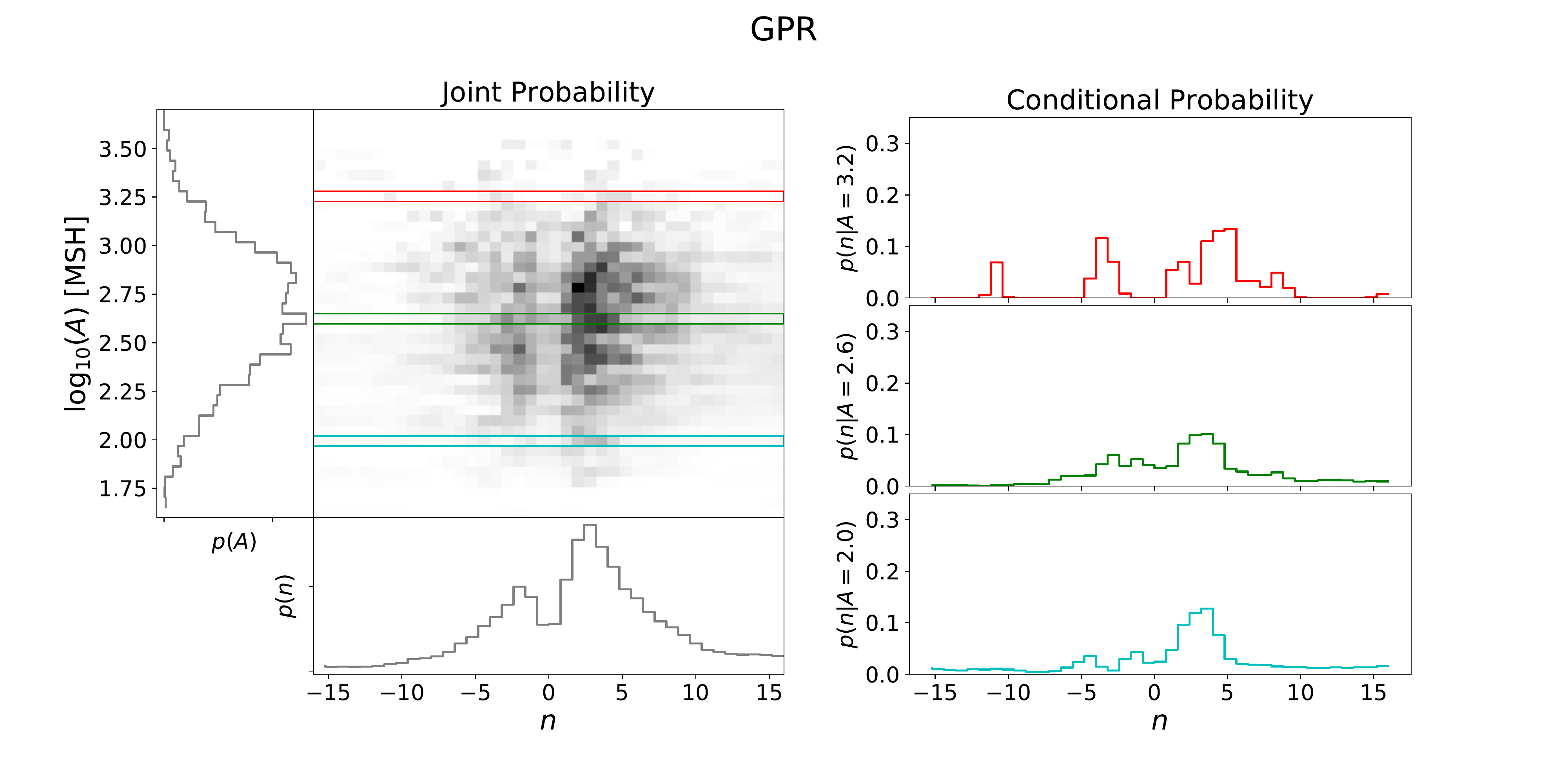}
    \includegraphics[width=0.49\linewidth]{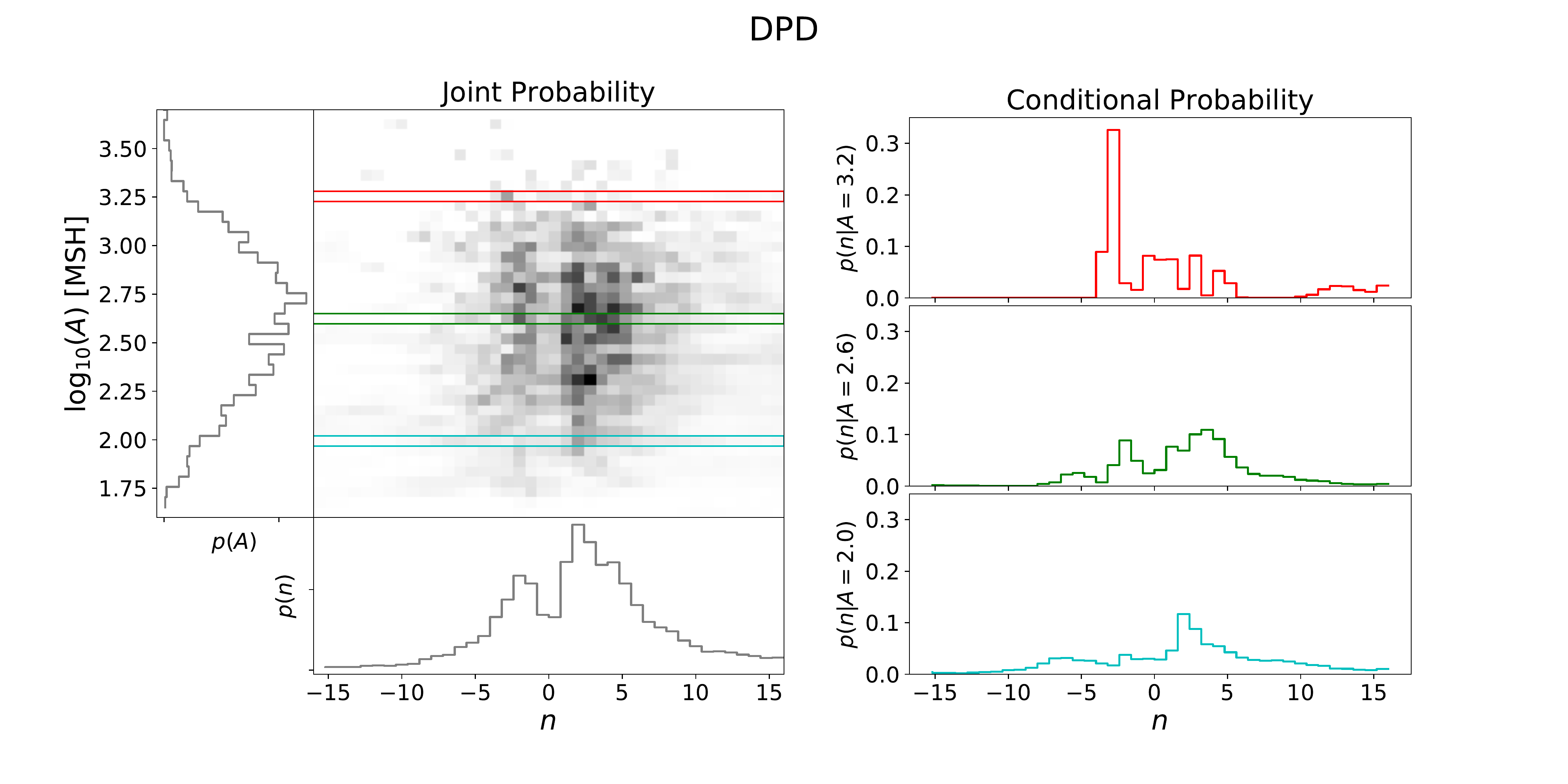}
    \caption{Joint probability density, $p(n, A)$, and the conditional probability density, $p(n \given A)$, for the GPR ({\it left-hand panel}) and DPD ({\it right-hand panel}) datasets. For each dataset the central left panel shows the joint probability (the grayscale corresponds to the value of the probability), the attached distributions show the marginal distributions of the area and the skew parameter (the grey colour spectrum is proportional to the magnitude of the probability). The right panels show the conditional probability at three different values of $A$ (in the figure showing the joint probability these three values correspond to  horizontal cuts in different colour).}
    \label{fig:prob-n-A}
\end{figure*} 

\begin{figure*}[]
    \centering
    \includegraphics[width=0.49\linewidth]{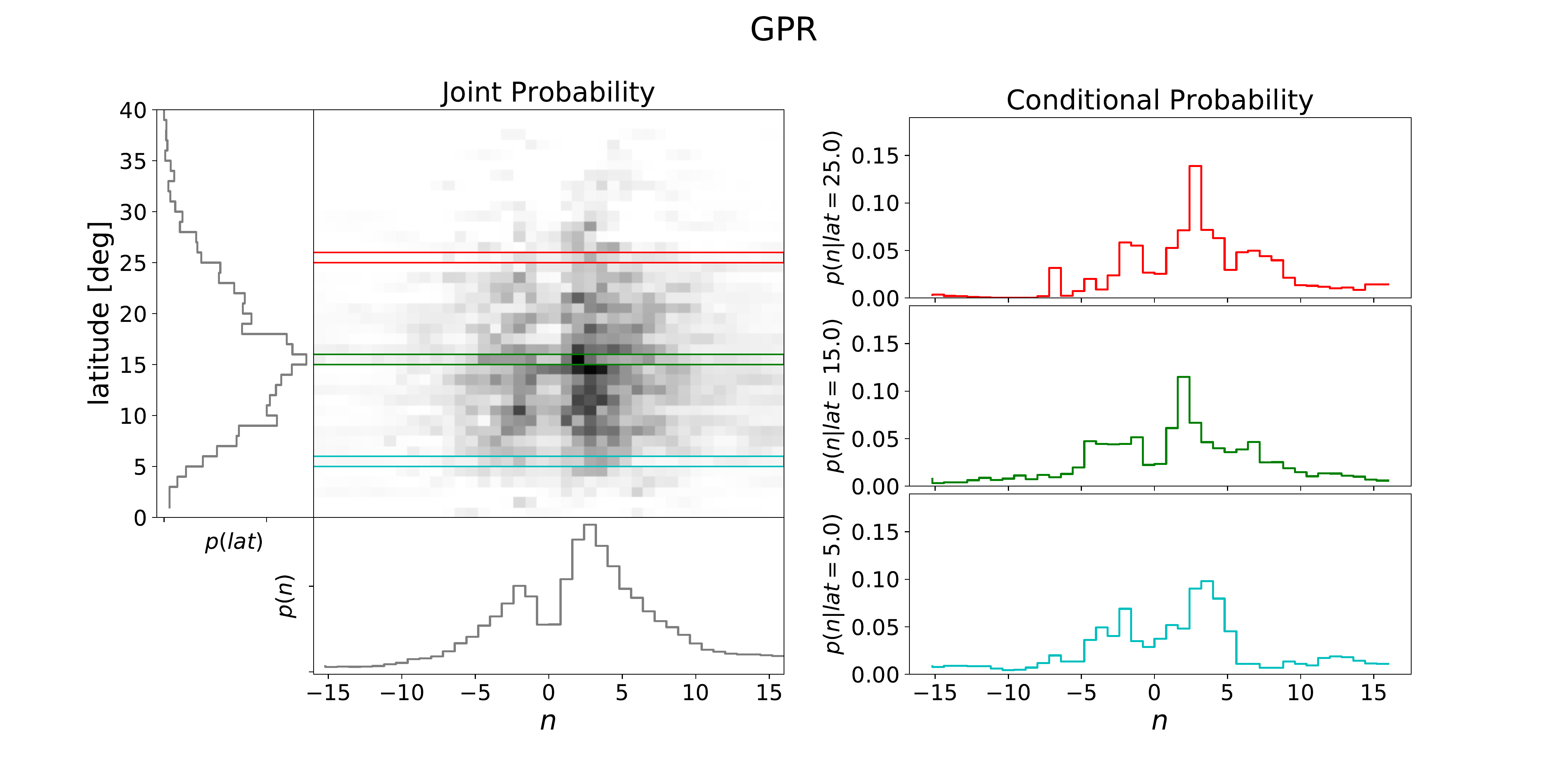}
    \includegraphics[width=0.49\linewidth]{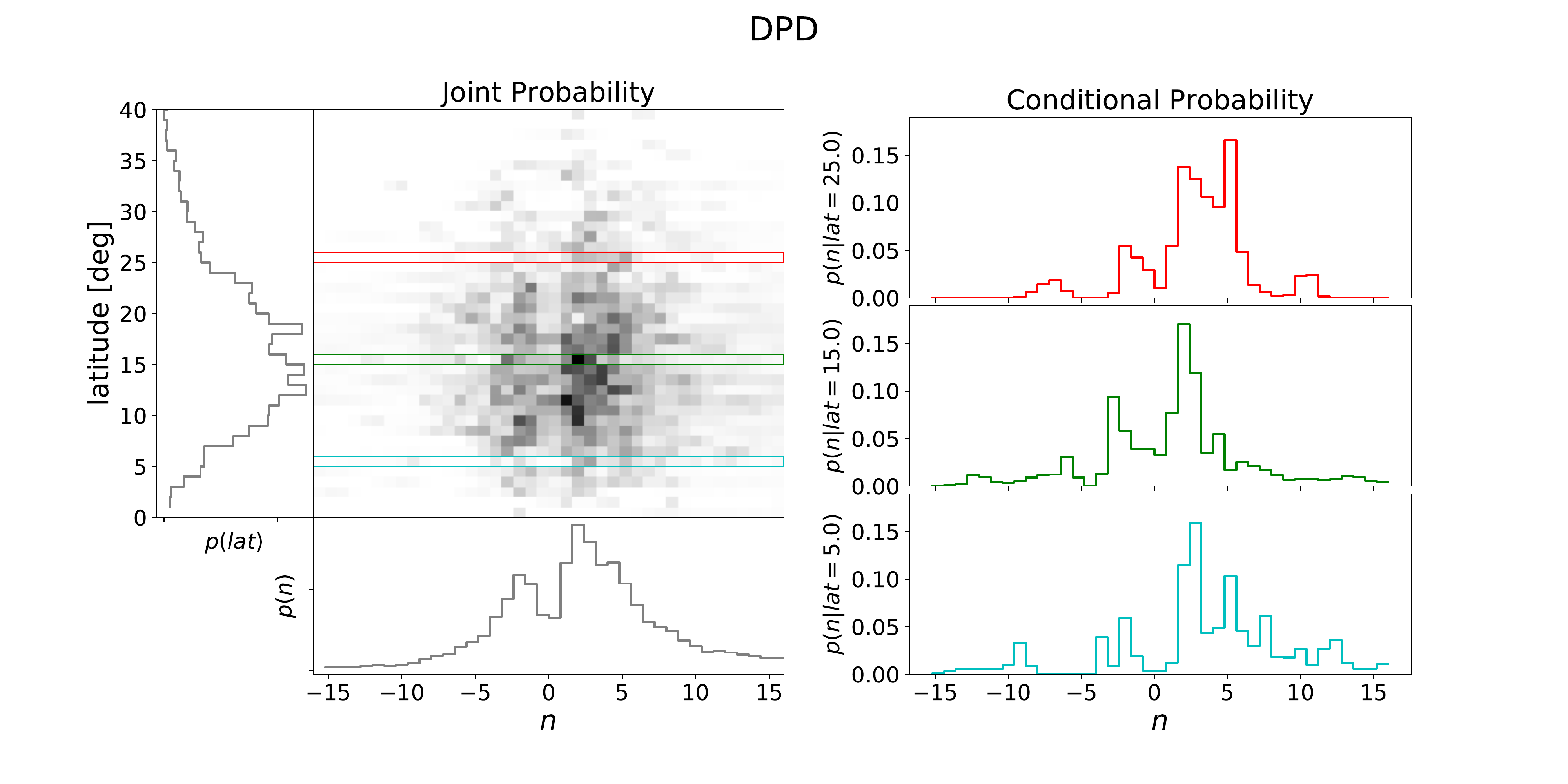}
    \caption{The same as in Fig.~\ref{fig:prob-n-A}, but here we plot the distribution of the skew parameter in terms of the latitude of the spot groups.}
    \label{fig:prob-n-lat}
\end{figure*} 

The normalised histograms of the lifetime of groups and the FWHM values derived from the fitted curves are shown in Fig.~\ref{fig:hist-lifetime}. In general the data obtained from the two databases are consistent (the lifetime of most of groups investigated by us is approximately 12 days), however, there are smaller differences between the two distributions. In the GPR data the groups with larger lifetime are more dominant than in the DPD database, while in the case of shorter lifetime groups is the opposite, that is, there are more groups with smaller lifetime in the DPD data. This finding requires more investigation but it is likely due to observational and data reduction effects. While sunspots in the groups in the GPR database were derived from solar photographs taken at various contributing solar observatories, which used different telescopes, experienced different seeing conditions, and employed different photographic processes, the DPD data is based on scanned photographic plates and CCD images reduced by software \citep[see e.g.][]{2017MNRAS.465.1259G}. These difference can very easily lead to strong positive or negative bias in area measurements that result in differences of the lifetime. Interestingly, both the FWHM and threshold-based lifetimes show the same effect which excludes the possibility that the differences arise solely from biased measurements of the small-area spots. We should mention that in the distribution of the lifetime (like in the case of the groups' area) we do not see the existence of two different types of sunspot groups.

Finally let us investigate the very interesting connectivity between the lifetime of sunspot groups and their total area. As we mentioned earlier, the most accepted relationship between these two parameters is the Gnevyshev-Waldmeier rule. In order to establish the relationship between the lifetime of groups and their area we plot the lifetime of groups in terms of their area in Fig.~\ref{fig:a-lt-fit}. In order to evidence even small changes we plot the logarithm of these values (left panel), however the deviation from a linear relationship is more visible in a linear scale diagram (right panel). On both panels we show the straight line corresponding to the well-known Gnevyshev-Waldmeier rule (green dashed line). In addition, we fitted our data with a straight line in the form of $T = \mathrm{const} \cdot A_\mathrm{max}$, where the parameter of the best fit is $\mathrm{const}=0.05\pm 0.006$, which is different from the previously established value of $0.1$. Since the scatter of points is far too large for a linear relationship, we concluded that it is much more appropriate look for a lower limit on the life-time of sunspot groups as a function of area. This envelop curve was obtained by fitting a straight line in the left panel to the data shown by black triangles, which on the right panel corresponds to a power law with an exponent less than one.

The equation of the curve determined this way is $T_{min} = (1.25 \pm 0.054) \cdot A_\mathrm{max}^{(0.33 \pm 0.021)}$. As a consequence, the minimum values of group lifetimes is approximately proportional to the cubic root of maximum area, i.e. $T_\mathrm{min} \propto \sqrt[3]{A_\mathrm{max}}$. However, we should note that a qualitative  description of the wide distribution of values is very volatile. Clearly the two  databases show a different scattering, therefore, we have repeated the same analysis for the two datasets separately (see Appendix A). The only important difference is the fitting parameters of the envelope curve, however the two sets of values are the same within the errors. The only plausible explanation for this difference is the way how these data have been recorded. Currently we have no physical basis for the existence of the envelope curves, our aim was to highlight this interesting behaviour. Surprisingly, the equation of the straight line of best linear fit is identical.

\subsection{Asymmetry of groups area temporal profiles}\label{subsec:asymmetry}

As we outlined in the Introduction, we are primarily interested in the skewed nature of sunspot group area curves, hypothesising that the slowly rising and rapidly decaying groups and the rapidly rising and slowly decaying groups form two distinct classes.

To evidence this hypothesis we plot the histogram of the skew parameter, $n$, in Fig.~\ref{fig:hist-n}. The histograms display 500 MCMC realizations for each of the spot groups for both data sets. The results clearly show that the distribution of the skew parameter is strongly bi-modal, and the peaks are centered on $n \sim -2.0$ and $n \sim 2.5$ are well-defined. The two databases are consistent with each other, but some important difference can be observed in the case of sunspot groups with a symmetric behaviour and the ones corresponding to negative skew-parameter, where the DPD database shows more examples of such groups than the GPR database. It is also obvious that the number of groups with positive skew parameter is larger than the number of groups with negative parameter, meaning that the sunspot groups corresponding to the rapidly rising and slowly decaying class are dominant. In addition, is it is also clear that the symmetric sunspot groups are just a small percentage of the total sample.

It is interesting to study the distribution of the skew parameter, $n$, as a function of the area of the spot groups. In Fig.~\ref{fig:prob-n-A} we plot the joint probability density, $p(n, A)$, and the conditional probability density, $p(n \given A)$, for a few different values of maximum area for the two datasets (more precisely the logarithm of groups' maximal area). The distributions are plotted using all Monte Carlo realizations of all spot groups of our samples. Based on these two figures it is evident that the bulk of the distributions are consistent, except some small differences. The distribution of the conditional probability shows that (i) in both datasets the small area groups have predominantly positive skew parameters; (ii) for intermediate size groups one can find both types (positive and negative skew parameters) and the distribution of groups is identical with the distribution pattern of the whole dataset, that is, the number of groups with positive $n$ is larger, (iii) for groups with large area there is a discrepancy between the two datasets but, in general, we can say that large sunspot groups have mostly negative skew parameter, which is most obvious in the DPD dataset. The above properties are in line with the findings by \citet{1992SoPh..137...51H}, who stated that ''...larger groups decay at higher rates than they grow and smaller groups grow at higher rates than they decay'', where slow growth and fast decay correspond to a negative skew parameter regime, while the fast growing and slow decaying case correspond to the case of positive skew parameter.

Similar to the above analysis, we can investigate the relationship between the skew parameter of sunspot groups and the position of their appearance. To elucidate the connectivity between these two aspects, we plot the joint probability density, $p(n, \theta)$, and the conditional probability density, $p(n \given \theta)$, for a few different values of their latitude (taken as absolute value) for both datasets (Fig.~\ref{fig:prob-n-lat}).  The conditional probability shows that the distribution of the skew parameter for low and high values of the latitude is highly symmetrical so that the number of groups with positive and negative skew are similar. In the middle of the activity belt, where most of the spots are situated, we can see 
a distribution more similar to those of all groups, as in Fig.~\ref{fig:hist-n}.


\section{Conclusions}\label{sec:conclusion}

The generation, evolution and decay of sunspots are some of the most important aspects of solar magnetism. The complexity of this problem resides in the multitude of mechanisms that are involved in these processes, starting from magneto-convection, flux emergence, magnetic erosion, evolution of active regions, etc. In our study we aimed to describe statistically the temporal evolution of the total area of sunspot groups. To substantiate our investigations, we used two, publicly available, databases (GPR and DPD) that provide observations on the temporal evolution of the total group area. Thanks to the large time-span of these historical observations, we constructed a database consisting of almost 4000 events. One of the novelties applied in our analysis was the use of Bayesian technique to fit a model to determine the temporal evolution of groups area. 

With the help of many Monte-Carlo realizations of the fitted curves that increased the stability of the method, we were able to reveal several important aspects and properties of sunspots. A great advantage of the technique developed in the present study is that it allows the consideration of even those sunspot groups whose initial growth and/or final decay phases are not available due to missing data, solar rotation, etc. Although the used databases do not contain observational errors -- meaning that we cannot compute real errors to the fitting -- the fitting errors were automatically taken into account when constructing the histograms from all MCMC realisation, rather than the best fit. 

Our results show that sunspot groups fall into two distinct categories, according to their skew parameter, $n$. Positive and negative values of $n$ describe the nature of the flux emergence and decay of sunspot groups and these appear to correlate with maximum area, with small/large sunspot groups tending to have a negative/positive skew parameter. When investigating the relationship of the groups' maximal area with their lifetime we could not find a clear relationship similar to the one found in earlier studies, instead, we defined a lower limit of the lifetime that varies proportional to the cubic root of the total area. In a subsequent paper we will discuss the evolution of the growth and decay rate of sunspot groups, with great implications to the study of flux emergence and erosive decay of sunspots.   

\begin{acknowledgements}
EF-D thanks the support of the Hungarian National Research, Development and Innovation Office (NKFIH), under the grant K-128384 and by the European Union’s Horizon 2020 research and innovation program under grant agreement No. 95562. We acknowledge the computational resources for the Wigner GPU Laboratory of the Wigner Research Centre for Physics. 

\end{acknowledgements}

%
   \bibliographystyle{aa} 
   \bibliography{main_v20200105} 
%

\begin{appendix}
\section{Lifetime scatter in terms of the area of sunspot groups for the two databases}

\begin{figure}[!h]
    \centering
    \includegraphics[width=1.1\linewidth]{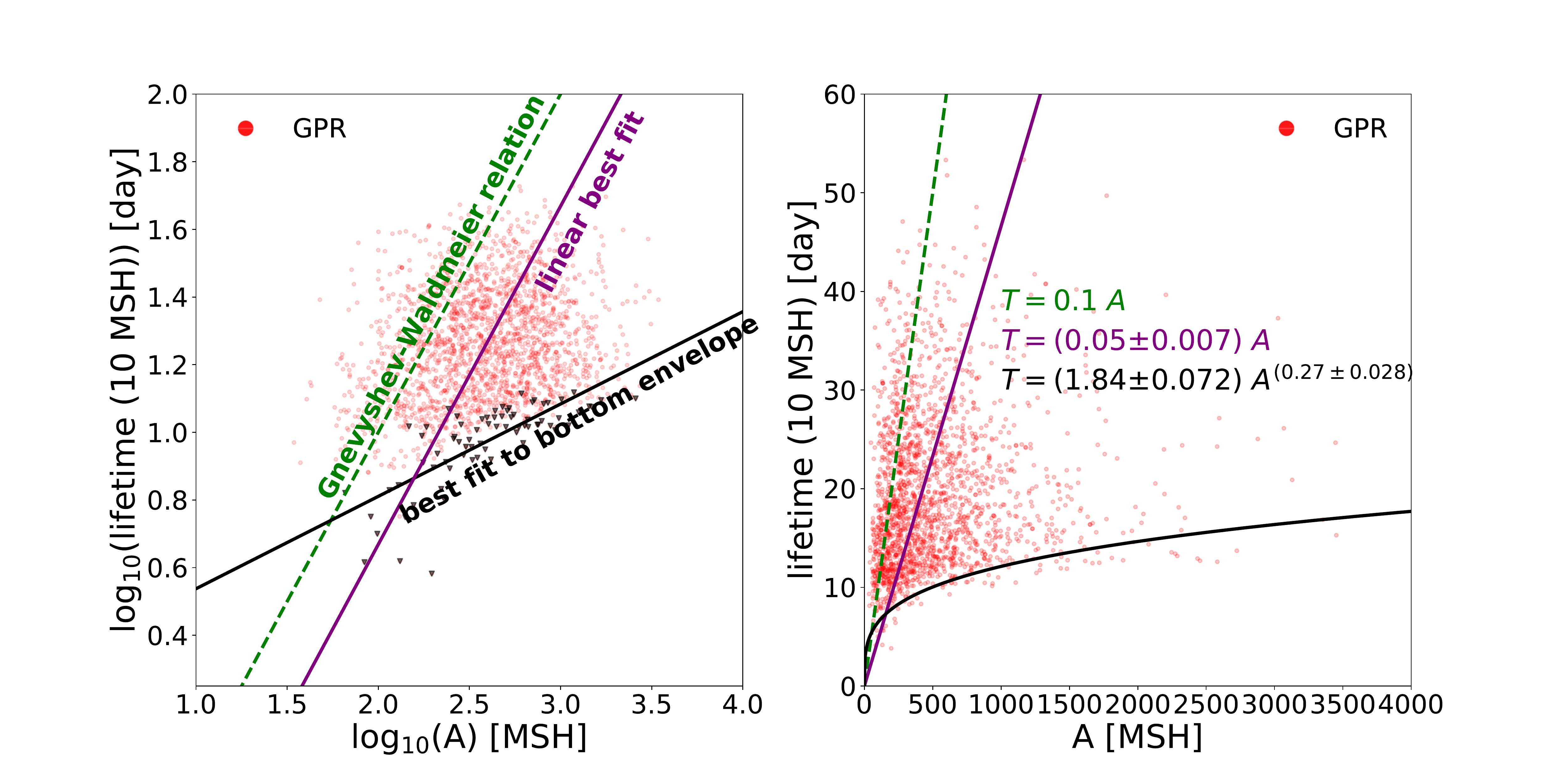}
    \caption{The lifetime of sunspot groups in terms of their area for the GPR data. In the left panel we plot the logarithmic values of the lifetime and the area, while in the right panel we plot their linear values. The dashed green line denotes the variation that corresponds to the Gnevyshev-Waldmeier rule, the purple solid line shows the best linear fit based on our data, while the black solid curve represents the lower envelop. The envelop curve has been obtained by fitting the groups shown by dark triangles. The errors shown in the equation of curves were derived from the standard deviation values. 
    }
    \label{fig:gpr-a-lt-fit}
\end{figure} 

\begin{figure}[!h]
    \centering
    \includegraphics[width=1.1\linewidth]{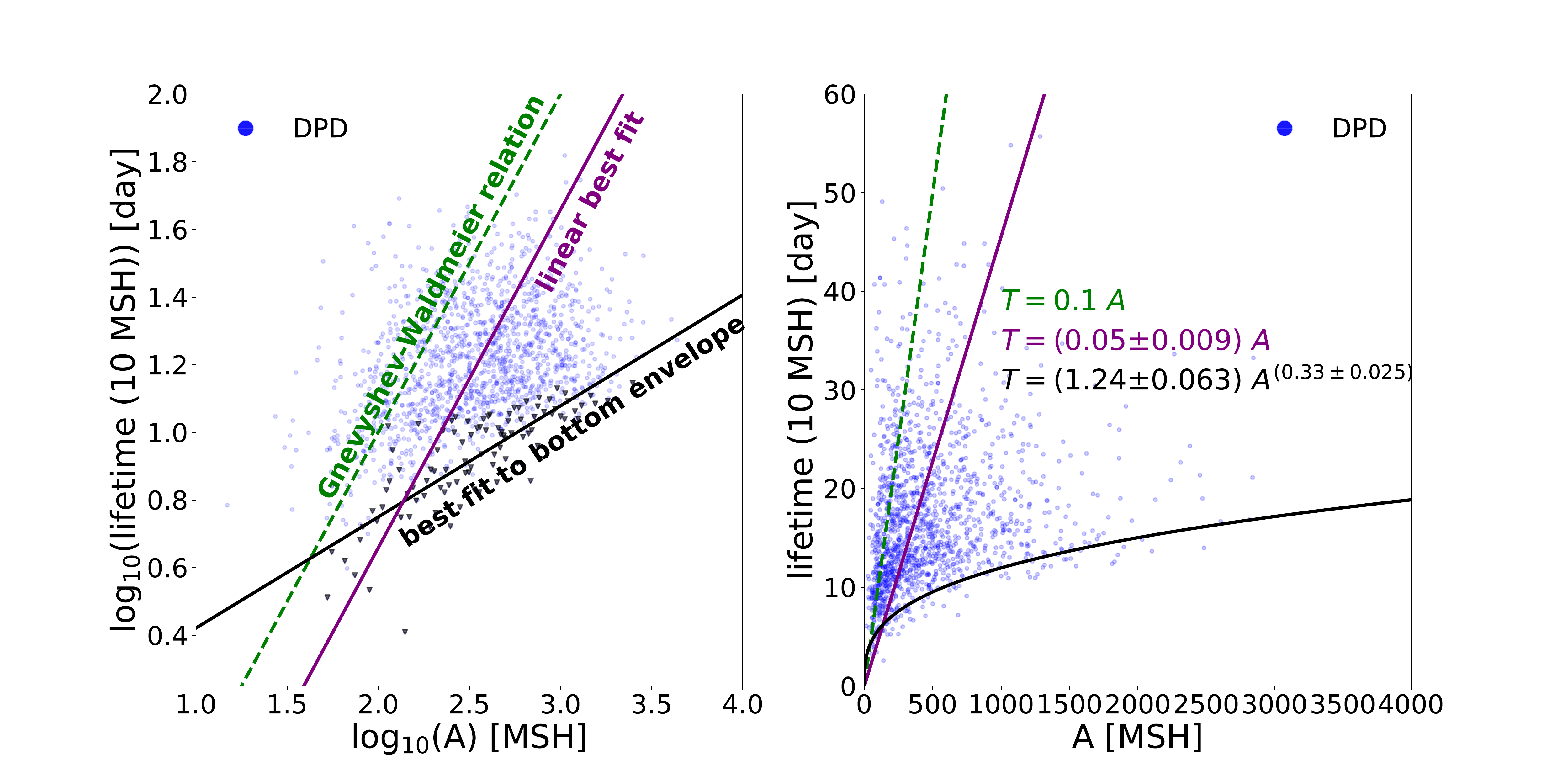}
    \caption{The same as Fig.~\ref{fig:gpr-a-lt-fit}, but for the DPD data.}
    \label{fig:dpd-a-lt-fit}
\end{figure} 

\end{appendix}

\end{document}